\documentclass{aa}

\usepackage{graphicx}
\usepackage{natbib}
\usepackage{booktabs}
\usepackage{txfonts}

\bibpunct{(}{)}{;}{a}{}{,} % to follow the A&A style

\def\beq{\begin{equation}}
\def\eeq{\end{equation}}
\def\beqn{\begin{eqnarray}}
\def\eeqn{\end{eqnarray}}

\begin{document}

\title{Uncertainties in the pasta-phase properties of catalysed neutron stars  
}
\titlerunning{Uncertainties in the pasta-phase properties}
\authorrunning{Dinh Thi et al.}

\author{H. Dinh Thi\inst{1},  
T. Carreau\inst{1},
A.~F. Fantina\inst{2},
F. Gulminelli\inst{1}}

\institute{Normandie Univ., ENSICAEN, UNICAEN, CNRS/IN2P3, LPC Caen, F-14000 Caen, France \\
  \email{dinh@lpccaen.in2p3.fr}
\and Grand Acc\'el\'erateur National d'Ions Lourds (GANIL), CEA/DRF - CNRS/IN2P3, Boulevard Henri Becquerel, 14076 Caen, France 
}
   
\date{Received xxx Accepted xxx}

\abstract{%Context. 
The interior of a neutron star is expected to exhibit different states of matter. In particular, complex non-spherical configurations known as `pasta' phases may exist at the highest densities in the inner crust, potentially having an impact on different neutron-star phenomena.}
{%Aim.
We study the properties of the pasta phase and the uncertainties in the pasta observables which are due to our incomplete knowledge of the nuclear energy functional.} 
{%Method.
To this aim, we employed a compressible liquid-drop model approach with surface parameters optimised either on experimental nuclear masses or theoretical calculations.
To assess the model uncertainties, we performed a Bayesian analysis by largely varying the model parameters using uniform priors, and generating posterior distributions with filters accounting for both our present low-density nuclear physics knowledge and high-density neutron-star physics constraints.
}
{%Results.
Our results show that the nuclear physics constraints, such as the neutron-matter equation of state at very low density and the experimental mass measurements, are crucial in determining the crustal and pasta observables.
Accounting for all constraints, we demonstrate that the presence of pasta phases is robustly predicted in an important fraction of the inner crust. We estimate the relative crustal thickness associated with pasta phases as $R_{\rm pasta}/R_{\rm crust}=0.128\pm 0.047$ and the relative moment of inertia as $I_{\rm pasta}/I_{\rm crust}=0.480\pm 0.137$. 
}
{%Conclusion.
Our findings indicate that the surface and curvature parameters are more influential than the bulk parameters for the description of the pasta observables.
We also show that using a surface tension that is inconsistent with the bulk functional leads to an underestimation of both the average values and the uncertainties in the pasta properties, thus highlighting the importance of a consistent calculation of the nuclear functional.
} 


\maketitle

%%%%%%%%%
\section{Introduction}
\label{sect:introd}
%%%%%%%%%

The interior of a cold neutron star (NS) is predicted to be made of various phases of matter, from a solid crust to a liquid core.
In particular, the outer crust is thought to be made of ions arranged in a lattice, embedded in an electron gas, while in the inner crust the neutron-proton clusters, neutralised by the electron gas, also coexist with a (free) neutron gas, until, at about half the saturation density ($n_{\rm sat} \approx 0.15$~fm$^{-3}$), nuclei dissolve into homogeneous matter, thus marking the transition to the liquid core \citep{hpy2007}.
In a liquid-drop picture of the nucleus, the equilibrium structure of the clusters in the crust results from the competition between the Coulomb and surface energy of nuclei.
At lower densities, in the outer crust and in most regions of the inner crust, nuclear clusters are sufficiently far apart from each other that their structure is not influenced by neighbouring clusters.
The effect of the surface energy prevails and clusters are thus expected to be spherical.
However, at a higher density, nuclei become very close to each other, and, eventually, at the bottom of the crust, matter may arrange itself into various exotic configurations known as `pasta' phases.
Although so far there is no direct observational evidence of pasta phases, their existence may have a sizeable impact on different NS phenomena, such as NS cooling  \citep{Newton2013b, hor2015,lin2020}, magnetic and rotational evolution of pulsars \citep{Pons2013}, crust oscillations \citep{gearheart2011, sotani2012}, and transport properties (see e.g. \citet{SchSht2018} for a review).
In addition, \citet{gearheart2011} also estimated the effect of non-spherical configurations on the NS `mountains', showing that the presence of pasta phases can decrease by up to an order of magnitude the NS maximum quadrupole ellipticity sustainable by the crust.

Since the pioneering works by \citet{Ravenhall1983}, \citet{Hashimoto1984}, and \citet{oya1984}, several studies have been conducted on the nuclear pasta, using different approaches, including compressible liquid-drop (CLD) models, the (extended) Thomas-Fermi method, nuclear energy-density functional theory, and molecular dynamics calculations (see \citet{Pethick1995, ch2008, wm2012, Blaschke2018} for a review and references therein).
Very recently, \citet{Balliet2020} studied the pasta properties in NSs within a CLD model with surface parameters fitted on quantum 3D Hartree-Fock calculations of nuclei in a neutron gas, and calculated the distributions of pasta observables, namely the mass and the thickness of the pasta layer, using a Bayesian analysis.
For all models they explored, they found that more than 50\% of the crust by mass and 15\% by thickness are made up of pasta, in agreement with the results of previous investigations \citep{LRP1993, Newton2013}.
Nevertheless, the presence of these non-spherical structures, such as cylindrical clusters (rods), slabs, cylindrical holes (tubes) and spherical holes (bubbles) in the NS, and the transition density among the different geometries, remain model dependent.

In this work, we studied the properties of the pasta phases in cold, non-accreting NSs under the so-called `catalysed matter hypothesis', that is matter in its absolute ground state at zero temperature.
To this aim, we extended the CLD model of \citet{Carreau2019,Carreau2020} to account for non-spherical pasta structures in the inner crust. 
This formalism is described in Sect.~\ref{sect:framework}.
The use of the CLD approach allows us to address the question of the model dependence of the results, which is discussed in Sect.~\ref{sect:models-cf} considering several representative nuclear functionals.
In addition, to obtain quantitative predictions of the uncertainties in the pasta observables, we performed a Bayesian analysis using flat uniform priors to generate posteriors with filters accounting for both our current knowledge of low-density nuclear physics and constraints at high density from general and NS physics.
This investigation is assessed in Sect.~\ref{sect:bayes}.
Finally, we present our conclusions in Sect.~\ref{sect:conclus}.

%%%%%%%%% 
\section{Theoretical framework}
%%%%%%%%%
\label{sect:framework}

%%%%%%%%%
\subsection{Model of the pasta phases}
\label{sect:model}
%%%%%%%%

Following the seminal work by \citet{bbp}, the equilibrium configuration of inhomogeneous catalysed matter (we assume zero temperature throughout the paper) present in the inner crust of NSs is variationally obtained within a CLD approach, along the same lines as in \citet{Douchin2001}, \citet{Carreau2019}, and \citet{Carreau2020}. 
We consider a periodic lattice configuration consisting of Wigner-Seitz cells of volume $V_{\rm WS}$ containing a clustered structure  (`pasta') %of volume $V$, 
composed of $Z$ protons of mass $m_p$ and $A-Z$ neutrons of mass $m_n$ ($A$ being the cluster total mass number), immersed in a uniform gas of neutrons and electrons with respective densities $n_g$ and $n_e$.
If inhomogeneities appear as clusters (holes) of volume $V$, the density distribution in the Wigner-Seitz cell is $n_i$ ($n_g$) if $l<r_N$, and $n_g$ ($n_i$) otherwise, $r_N$ being the linear dimension of the pasta structure and $l$ the linear coordinate of the Wigner-Seitz cell.
The denser clustered phase is characterised by a density $n_i=A/V$ ($n_i=A/(V_{\rm WS}-V)$ in the case of holes)
and a proton fraction $y_p=Z/A=n_p/n_i$. 
The volume fraction occupied by the cluster (of density $n_i$) or hole (of density $n_g$), $u=V/V_{\rm WS}$, thus reads
\begin{equation}
u=  \left \{
   \begin{array}{r c l}
      (n_B-n_g)/(n_i-n_g)  &  & \mbox{for clusters, and}  \\
     (n_i-n_B)/(n_i-n_g)  &  & \mbox{for holes}.
    \end{array}
   \right . 
\end{equation}

The energy density of the cell is minimised with the constraint of a given total baryonic density $n_B=n_n +n_p$, $n_p$ ($n_n$) being the proton (neutron) density, respectively.
Moreover, charge neutrality holds, thus $n_e=n_p$.
The corresponding thermodynamic potential per unit volume is written as 
\begin{eqnarray}
\Omega
&=& n_n m_n c^2+n_p m_p c^2 + \epsilon_B(n_i,1-2y_p) f(u) \nonumber \\ 
&+& \epsilon_B(n_g,1) (1-f(u)) + \epsilon_{\rm Coul} + \epsilon_{\rm {surf+curv}} \nonumber \\
&+& \epsilon_e -\mu_B^{\rm tot} \ n_B , 
\label{eq:auxiliary}
\end{eqnarray}
where $\epsilon_B(n,\delta)$ is the energy density of uniform nuclear matter at baryonic density $n$ and isospin asymmetry $\delta=(n_n-n_p)/n$, $\epsilon_e$ is the energy density of a pure uniform electron gas at density $n_e=n_p$, $\mu_B^{\rm tot}$ is the baryonic chemical potential (including the rest mass), 
and $\epsilon_{\rm {surf+curv}}$ and $\epsilon_{\rm Coul}$ are the finite-size corrections accounting for the interface tension between the cluster and the neutron gas and the electrostatic energy density, respectively. 
Finally, the function $f(u)$ is given by 
\begin{equation}
f(u)=  \left \{
   \begin{array}{r c l}
      u  &  & \mbox{for clusters, and} \\
     1-u  &  & \mbox{for holes}.
    \end{array}
   \right . 
\end{equation}

As was recognised in early works (see e.g. \citet{Ravenhall1983,Hashimoto1984}), one of the advantages of the decomposition of Eq.~(\ref{eq:auxiliary}) in terms of bulk and interface is that the geometry of the pasta structures only affects the finite-size corrections, and the latter can be expressed analytically as a function of the dimensionality of the structure ($d=1$ for slabs, $d=2$ for cylinders, $d=3$ for spheres).
We write the interface energy density as in \citet{Maru2005} and \citet{Newton2013}:
\begin{equation}
\epsilon_{\rm {surf+curv}}=\frac{ud}{r_N}\left ( \sigma_s +\frac{(d-1)\sigma_c}{r_N}\right ) , \label{eq:interface}   
\end{equation}
where the surface tension $\sigma_s$ and curvature tension $\sigma_c$  are independent of the dimensionality.
For the latter quantities, we used the expressions originally proposed by \citet{Ravenhall1983} on the basis of Thomas-Fermi calculations at extreme isospin asymmetries, also subsequently employed in different works on NS crust and supernova modelling within the CLD approximation \citep{lattimer1991, LRP1993, Newton2013,Carreau2019,Carreau2020,Balliet2020}, namely,
\begin{eqnarray}
\sigma_s&=&\sigma_0\frac{2^{p+1}+b_s}{y_p^{-p}+b_s+(1-y_p)^{-p}} \ , \label{eq:surface} \\
\sigma_c&=&5.5 \, \sigma_s \frac{\sigma_{0,c}}{\sigma_0}(\beta-y_p)\ , \label{eq:curvature} \ 
\end{eqnarray}
where the parameters $(\sigma_0,\sigma_{0,c},b_s,\beta,p)$ must be optimised on theoretical calculations or experimental data. 
The Coulomb energy density reads as follows:
\begin{equation}
\epsilon_{\rm Coul}= 2\pi \left ( e y_p n_i r_N \right )^2 u \eta_d , \label{eq:coulomb}   
\end{equation}
with $e$ being the elementary charge, and
\begin{eqnarray}
\eta_1&=& \frac{1}{3}\left [ u- 2 \left ( 1- \frac{1}{2u}\right )  \right ],  \label{eq:eta1} \\
\eta_2&=& \frac{1}{4}\left [ u - \ln u -1\right ],  \label{eq:eta2} \\
\eta_3&=&\frac{1}{5}\left [ u+ 2 \left ( 1- \frac{3}{2}u^{1/3} \right ) \right ]. \label{eq:eta3}
\end{eqnarray}

A nuclear model consists in a choice for the energy functional $\epsilon_B(n,\delta)$, complemented by a set of values for the surface parameters $(\sigma_0,\sigma_{0,c},b_s,\beta,p)$. 
This point is detailed in the next sub-section. 
Once the nuclear model is specified, the pasta structure and composition at a given baryonic density $n_B$  is determined by a two-step process. 
First, a geometry $d$ and a shape $s$ (clusters or holes) is considered, and the thermodynamical potential of Eq.~(\ref{eq:auxiliary}) is minimised with respect to the variational parameters $(n_i,I=1-2y_p, A,n_p,n_g)$. 
This allows us to identify the baryonic chemical potential $\mu = \mu_B^{\rm tot} - m_n c^2$ with the chemical potential of the neutron gas,
\begin{equation}
\mu=\frac{d\epsilon_B(n_g,1)}{dn_g} \ ,
\end{equation}
and gives the optimal value of $\Omega=\Omega_{\rm opt}(d,s)$ for each geometry. 
Then, the equilibrium configuration is defined as the one corresponding to the values of $d$ and $s$ that produce the minimum value of $\Omega_{\rm opt}$.

%%%%%%%%%
\subsection{Energy functional}
\label{sect:functional}
%%%%%%%%

The energy density $\epsilon_B$ of homogeneous nuclear matter is poorly known out of the saturation point $n_{\rm sat}\approx 0.15$ fm$^{-3}$ of symmetric $n_n=n_p$ matter, and a large number of models have been proposed in the literature based on effective Hamiltonians or Lagrangians. 
\citet{Margueron2018a} showed that those different models can be accurately reproduced using a Taylor expansion in $x=(n-n_{\rm sat})/3n_{\rm sat}$ up to order $4$ around the saturation point $(n=n_{\rm sat},\delta=0)$, varying the parameters of the expansion, which correspond to the so-called equation-of-state empirical parameters:
\begin{equation}
\epsilon_B(n,\delta) \approx n \sum_{m=0}^4 \frac{1}{m!} \left ( \left. \frac{d^m e_{\rm sat}}{d x^m} \right|_{x=0} 
+ \left.  \frac{d^m e_{\rm sym}}{d x^m} \right|_{x=0}\delta^2\right ) x^m .
\label{eq:etotisiv}
\end{equation} 
In Eq.~(\ref{eq:etotisiv}), $e_{\rm sat}=\epsilon_B(n,0) / n$ is the energy per baryon of symmetric matter, and $e_{\rm sym}=(\epsilon_B(n,1)-\epsilon_B(n,0))/ n$ is the symmetry energy per baryon, defined here as the difference between the energy of pure neutron matter and that of symmetric matter. 
To speed up the series convergence, a $\delta^{5/3}$ term from the fermionic zero-point energy is added, as well as an exponential correction ensuring the correct limiting behaviour at zero density (see Eq.~(17) in \citet{Carreau2019}). 
Following the common notation in the literature, we denote $E_{\rm sat(sym)}=e_{\rm sat(sym)}(n=n_{\rm sat})$, while the successive derivatives of $e_q$, with $q=$~sat,sym, are called $L_q,K_q,Q_q,Z_q$. 
The bulk parameters $\{E_q,L_q,K_q,Q_q,Z_q, q=$~sat,sym$\}$ are complemented with the saturation density parameter, $n_{\rm sat}$, two parameters related to the isoscalar effective mass and effective mass splitting, which are denoted as $\mathcal{K}_q$, and the $b$ parameter governing the functional behaviour close to the zero-density limit (see Sect.~2.2 in \citet{Carreau2019} for details).
The complete parameter set thus has 13 parameters and will be noted in a compact form as $\vec X_{\rm bulk}\equiv\{n_{\rm sat}, b, (E_q,L_q,K_q,Q_q,Z_q, \mathcal{K}_q, q=$~sat,sym$)\}$.
Different nuclear models will then correspond to different sets of $\vec X_{\rm bulk}$ parameters.
 
Concerning the parameters $(\sigma_0,\sigma_{0,c},b_s,\beta,p)$ entering the surface functional, Eq.~(\ref{eq:interface}), 
 \citet{Carreau2019} showed that, in order to have a realistic and consistent treatment of the composition of the crust, it is important that these parameters are fixed consistently with the functional employed for the bulk part of the nuclear energy $\epsilon_B$. 
 Indeed, the bulk energy is model dependent, but in the case of a spherical geometry the total (bulk plus surface plus Coulomb) energy is constrained by the requirement of reproducing the experimentally measured nuclear masses, which creates an obvious correlation between the bulk and surface parameters that should be accounted for.

In the vacuum, the nuclear mass corresponding to a spherical fully ionised atom of charge $Z$ and mass number $A$ can be deduced from Eqs.~(\ref{eq:auxiliary}), (\ref{eq:interface}), and (\ref{eq:coulomb}) as:
\begin{eqnarray}
M(A,Z) c^2&=&m_pc^2Z+m_nc^2(A-Z) \nonumber \\ 
&+& \frac{A}{n_0}  \epsilon_B(n_0,I) +4\pi r_N^2\left (\sigma_s +\frac{2\sigma_c}{r_N}\right ) \nonumber \\ 
&+& \frac{3}{5}\frac{e^2Z^2}{r_N} , \label{eq:mass}
\end{eqnarray} 
where $I=1-2Z/A$, the nuclear radius is $r_N=(4\pi n_0/3)^{-1/3} A^{1/3}$, and the bulk density $n_0$  is given by the equilibrium density of nuclear matter at isospin asymmetry $I$, defined by $\partial \epsilon_B/\partial n|_{I,n_0}=0$.

For each choice of the parameter set $\vec X_{\rm bulk}$, we determine the associated surface parameters by a $\chi^2$-fit of  Eq.~(\ref{eq:mass}) to the experimental Atomic Mass Evaluation (AME) 2016 \citep{AME2016}.

As an example, in Table~\ref{tab:bulkpar} we give the bulk parameters that reproduce the density behaviour of symmetric nuclear matter and pure neutron matter of a chosen set of popular nuclear models that are analysed in the next section.
The associated optimal surface parameters are given in Table \ref{tab:surfpar}.
The ($\sigma_0$, $b_s$, $\sigma_{0,c}$, $\beta$) parameters were obtained from the fit of the experimentally measured nuclear masses \citep{AME2016}, while the $p$ parameter was optimised to provide a good reproduction of the crust–core transition density of the different functionals, whenever available, or fixed to $p=3$ otherwise (see Table~\ref{tab:literature} and \citet{Carreau2019, Carreau2020} for a discussion).

\begin{table*}[!htbp]
%\begin{center}
   \caption{Bulk parameters used in the Taylor expansion to reproduce the low-density behaviour of symmetric matter and pure neutron matter of different models.
   }
 \centering
  \begin{tabular}{lcccccccccc}
    \toprule
    & $n_{\rm sat}$ & $E_{\rm sat}$ & $E_{\rm sym}$ & $L_{\rm sym}$ & $K_{\rm sat}$ & $K_{\rm sym}$ & $Q_{\rm sat}$ & $Q_{\rm sym}$ & $Z_{\rm sat}$ & $Z_{\rm sym}$ \\
    & (fm$^{-3}$) & (MeV) & (MeV) & (MeV) & (MeV) & (MeV) & (MeV) & (MeV) & (MeV) & (MeV)   \\
    \midrule
    BSk24  &  0.1578 & -16.05  & 30.00  & 46.4  & 245.5  & -37.6   & -274.5  & 710.9    & 1184.2   &  -4031.3   \\ 
    SLy4    &  0.1595  & -15.97  & 32.01  & 46.0  & 230.0  & -120.0 & -363.0 & 521.0 & 1587.0   & -3197.0 \\
    RATP  &  0.1598  & -16.05  & 29.26  &  32.4 & 240.0  & -191.0  & -350.0   & 440.0   & 1452.0  & -2477.0  \\ 
    NRAPR &  0.1606  & -15.85  & 32.78  & 59.7 & 226.0 & -123.0 & -363.0 & 312.0 & 1611.0 & -1838.0 \\
    DD-ME2&  0.1520   & -16.14  & 32.31  & 51.3  & 251.0  & -87.0   &  479.0  & 777.0 & 4448.0 & -7048.0   \\
    DD-ME$\delta$&  0.1520   & -16.12  & 32.35  & 52.8  & 219.0  & -118.0   &  -748.0  & 846.0 & 3950.0 & -3545.0   \\
    NL3    &  0.1480    & -16.24  & 37.35  & 118.3 & 271.0  & 101.0  & 198.0 & 182.0 & 9302.0 & -3961.0 \\
    PKDD & 0.1495    & -16.27 & 31.19   & 79.5 & 261.0   & -50.0   & -119.0    & -28.0 & 4213.0 &  -1315.0 \\ 
    TM1 & 0.1450 & -16.26 & 36.94 & 111.0 & 281.0 & 34.0 & -285.0 & -67.0 & 2014.0 & -1546.0  \\
    \bottomrule
  \end{tabular}
  \tablefoot{Parameters listed in the tables, $\{n_{\rm sat}, E_q,L_q,K_q,Q_q,Z_q, q=$~sat,sym$\}$, are used to reproduce the effective models 
   BSk24 from \citet{BSK24}, 
   SLy4 from \citet{SLy4}, 
   RATP from \citet{RATP}, 
   NRAPR from \citet{NRAPR}, 
   DD-ME2 from \citet{DDME2}, 
   DD-ME$\delta$ from \citet{DDMEd}, 
   NL3 from \citet{NL3}, 
   PKDD from \citet{PKDD},
   and TM1 from \cite{TM1}.
   }
\label{tab:bulkpar}
%\end{center}
\end{table*}

\begin{table}[!htbp]
   \caption{Optminised surface and curvature parameters for different functionals, for which the bulk parameters are fixed as in Table~\ref{tab:bulkpar}.}
   \setlength{\tabcolsep}{3pt}
 \centering
  \begin{tabular}{lccccc}
    \toprule
    &$\sigma_0$&$b_s$&$\sigma_{0,c}$&$\beta$& $p$\\
    & (MeV/fm$^2$)&& (MeV/fm) &&\\
    \midrule
    BSk24               &1.05021&30.32168&0.12147&0.66495&3.00\\ 
    SLy4                &0.98911&19.02416&0.15141&0.75548&3.00\\
    RATP                &1.05161&35.22683&0.12488&0.67745&3.00\\ 
    NRAPR               &0.91932&14.68853&0.16594&0.83634&3.00\\
    DD-ME2              &1.09358&5.47648 &0.11969&0.53966&2.42\\  
   DD-ME$\delta$       &1.08385&11.38970&0.11810&0.55559&2.73\\
    NL3                 &1.12493&4.52517 &0.12297&0.46301&2.79\\
    PKDD                &1.17354&27.70134&0.08008&0.25816&3.00\\ 
    TM1                 &1.13817&9.31146 &0.11377&0.39118&3.00\\
    \bottomrule
  \end{tabular}
\label{tab:surfpar}
\end{table}
%
%

%%%%%%%%%
\section{Comparison of different nuclear functionals}
\label{sect:models-cf}
%%%%%%%%

\begin{figure}[htbp]
\begin{center}
\includegraphics[width=\linewidth]{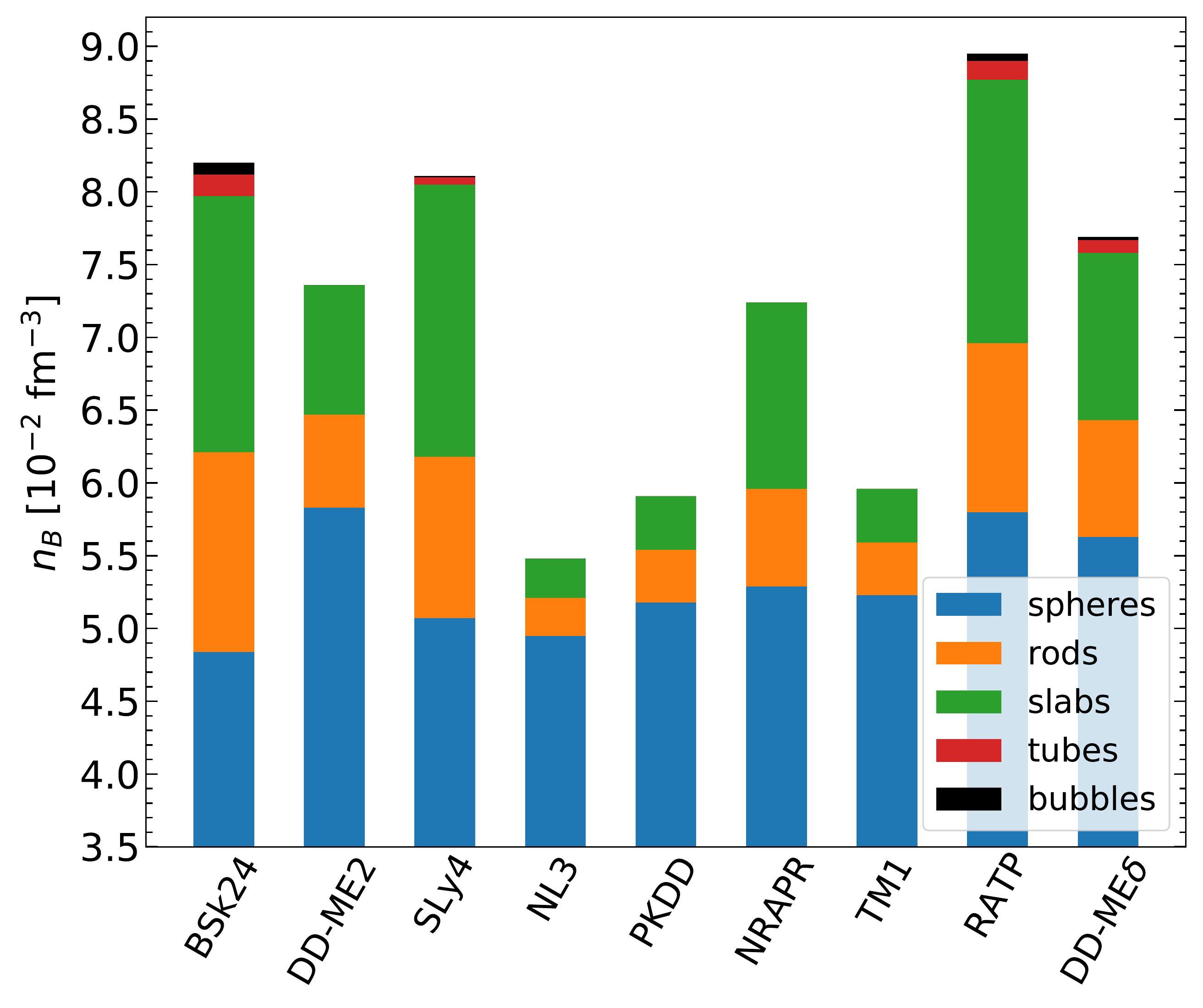}
\end{center}
\caption{Sequence of equilibrium pasta phases and corresponding transition densities among the geometries for different nuclear functionals
(see text for details). }
\label{fig:compo}
\end{figure}

Using the formalism described in Sect.~\ref{sect:framework}, we computed the properties of the pasta layer predicted at the bottom of the inner crust.
To explore the model dependence of the results, we examined the equilibrium configurations obtained with different popular nuclear models, corresponding to the different parameter sets listed in Tables \ref{tab:bulkpar} and \ref{tab:surfpar}, as illustrative examples.
Results are shown in Fig.~\ref{fig:compo}, where the different colours correspond to the density regions where different geometries (spheres, rods, slabs, tubes, and possibly bubbles) dominate. 
The upper edge of each column gives the transition point from the inhomogeneous crust to the homogeneous core, defined as the point where the thermodynamical potential, Eq.~(\ref{eq:auxiliary}), corresponding to the optimal geometry, equals the thermodynamical potential of homogeneous nuclear matter in beta equilibrium.
The sequence of the different geometries appears to be model independent, although not all the considered models predict bubble configurations, and it is consistent with previous results (see \citet{Pethick1995} for a review).

For a given energy functional, it is well known that the precise value of the transition density from the spherical to the non-spherical phases depends on the details of the many-body model used to treat the clustered structure. 
A comparison with results in the literature is reported in Table~\ref{tab:literature}. 
The most sophisticated calculation presently available is the one by \citet{Pearson2020} with full fourth-order extended Thomas-Fermi calculations within the BSk24 functional. 
In that paper, the authors obtained a core-crust transition density very close to the one displayed in Fig.~\ref{fig:compo} for BSk24 and a slightly higher value for the transition density to the rod shape (see Table~\ref{tab:literature}).
We consider this comparison as a very good agreement, especially considering that beyond-mean-field contributions such as pairing, which were fully neglected here, have recently been shown to affect the composition of the crust \citep{Pearson2020, Shelley2020}. 

\begin{table}[htbp]
        \caption{Transition density from spheres to cylinders ($n_{\rm p}$) and from crust to core ($n_{\rm CC}$) for different functionals.}
        \centering
        \setlength{\tabcolsep}{2pt}
        \renewcommand{\arraystretch}{1.2}
        \begin{tabular}{l|cccccc}
                \hline
                \multicolumn{1}{c}{} & \multicolumn{6}{c}{$n_{\rm p}/n_{\rm CC}$ ($10^{-2}$~fm$^{-3}$ )} \\
                Model &  This & Pearson  & M\&U & D\&H & Vi\~{n}as & Grill     \\
                           &work& et al. && & et al. & et al. \\
                \hline
             BSk24   &4.84/8.20 & 5.0/8.1 & &  &  &   \\
          SLy4 &5.07/8.11 && 6.1/8.1  & -/7.6 & -/7.6 &    \\
          NL3     &4.95/5.48 &&  &  &  & -/5.48  \\
          DD-ME2  &5.83/7.36 &&  &  &  & 6.11/7.35  \\  
          DD-ME$\delta$&5.63/7.69 && & & & 6.26/7.66\\
                \hline
        \end{tabular}
        \tablefoot{For comparison, results from \citet{Pearson2020}, \citet{Martin2015} (M\&U), \citet{Douchin2001} (D\&H), \citet{Vinas2017}, and \citet{Grill2012} are given. The `-' sign indicates that no transition to pasta is found.}
        \label{tab:literature}
\end{table}

In the case of the SLy4 functional, \citet{Martin2015} employed a slightly less sophisticated second-order extended Thomas-Fermi approach, and they observed a transition to the cylindrical shape at a higher density than the value obtained with our approach for SLy4.
However, the thermodynamical potentials corresponding to $d=3$ and $d=2$ in \citet{Martin2015} are almost indistinguishable starting from $n=0.05$ fm$^{-3}$, in good agreement with our results. 
The crust-core transition point is also in excellent agreement with our findings. 
At variance with this result, \citet{Vinas2017} and \citet{Douchin2001}, who also employed the SLy4 functional, reported no deviation from the spherical shape within a zeroth-order Thomas-Fermi calculation and a CLD approach, respectively.
Moreover, they also obtain a lower density for the crust-core transition.
Finally, we can also compare our results concerning the relativistic functionals NL3, DD-ME2, and DD-ME$\delta$ with the extensive Thomas-Fermi calculations of \citet{Grill2012}; in our approach, we obtain a lower transition density between spheres and cylinders, and we do find non-spherical configurations for the NL3 functional, unlike \citet{Grill2012}. 

Even if the comparison of our work with previous results can be globally considered as satisfactory, Table \ref{tab:literature} shows that some differences exist between the inner-crust composition of different many-body methods used to compute the energy of clusterised matter, for a fixed equation of state (or, equivalently, a fixed set of bulk parameters). 
This highlights the importance of finite-size contributions to the nuclear energy, which are essential in determining the optimal composition and are not uniquely linked to the bulk matter properties. 
Only the bulk parameters of our meta-model are adjusted to reproduce a given functional, while we employ a different fitting protocol for the surface (plus curvature) energy.
The effects of the surface contribution are studied in greater detail in Sect.~\ref{sect:surface}.

\begin{figure*}[!htbp]
\begin{center}
\includegraphics[width=0.8\linewidth]{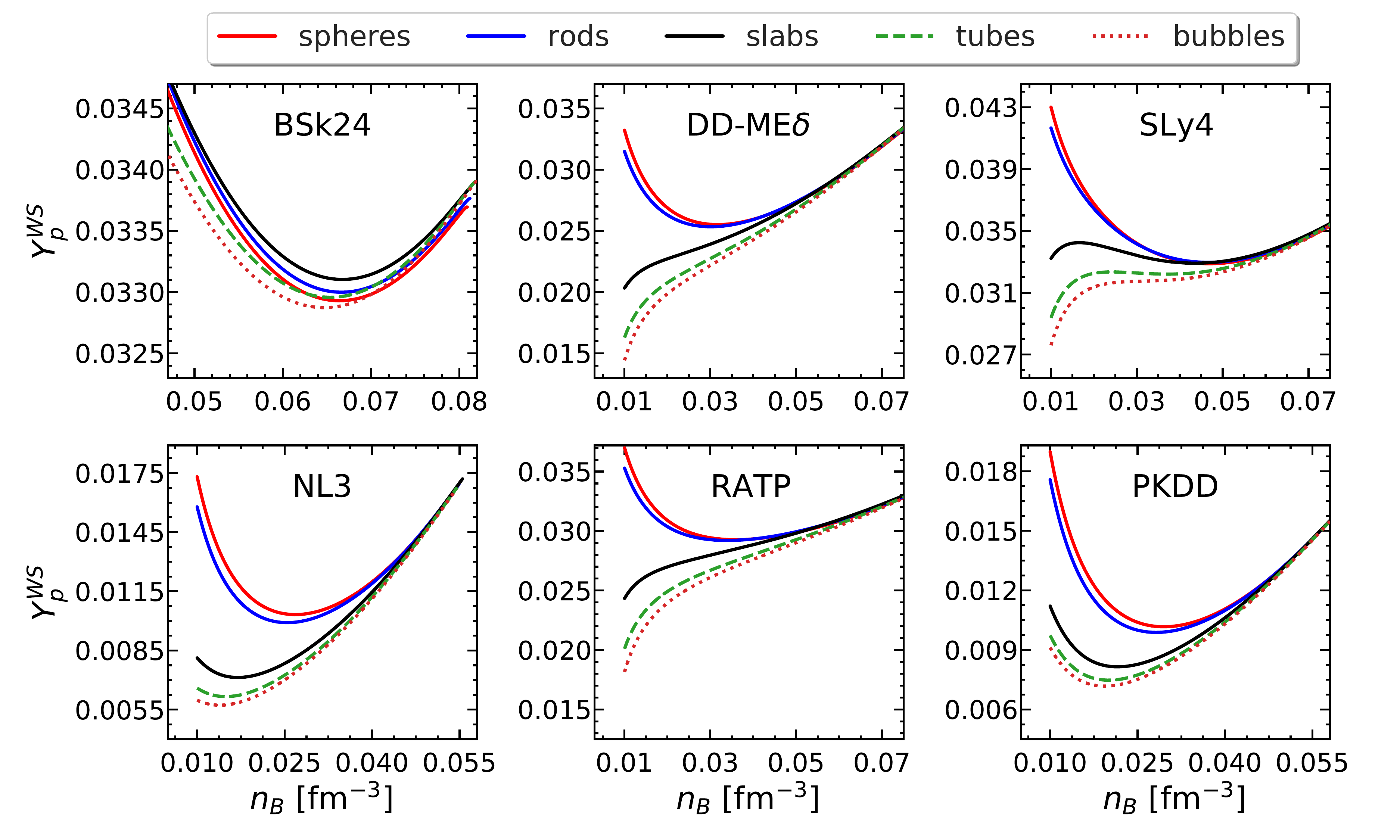}
\caption{Proton fraction in the Wigner-Seitz cell as a function of the baryonic density in the neutron-star crust for different geometries of the clustered structures using different popular nuclear models 
(see text for details).}
\label{fig:yp}
\end{center}
\end{figure*}

A complementary piece of information on the inner-crust composition in the presence of non-spherical geometries  is given 
by Fig.~\ref{fig:yp}, which reports the total proton fraction in the cell, $Y_p^{\rm WS}$, as a function of the baryonic density for the different geometries, for a selected set of functionals, as illustrative examples. 
We can see that some of the functionals exhibit the characteristic parabolic shape already reported by \citet{Pearson2020} for the BSk24 model, and the absolute value of the proton fraction is also in reasonable agreement with the findings of \citet{Pearson2020}. 
However, we can also observe that the trend of the proton fraction, the effect of the geometry, and the numerical value throughout the inner crust are very strongly model dependent.

To summarise  the previous discussion, from Figs.~\ref{fig:compo} and \ref{fig:yp} we can conclude that the uncertainty in the predictions due to the choice of the nuclear functional is much more important than the uncertainty due to the many-body method adopted to describe the inhomogeneities. 
We limited our analysis to a small set of popular models, that span a quite wide range of the empirical parameters, to provide illustrative examples.
However, these nine models might not fully encompass the current uncertainties in the nuclear equation of state.
For this reason, in the next section we assess the model dependence of the crust composition in a more quantitative way by largely exploring the parameter space of the nuclear functionals compatible with the present theoretical and experimental constraints.

%%%%%%%%%
\section{Bayesian analysis}
\label{sect:bayes}
%%%%%%%%

We now come to the quantitative determination of the uncertainty on the pasta observables, due to the incomplete knowledge of the nuclear energy functional. 
To this aim, we performed a Bayesian analysis by largely varying the model parameters $\vec X=\{ \vec{X}_{\rm bulk}, p \}$ using flat non-informative priors, and generating posterior distribution with filters that impose both our present low-density (LD) nuclear physics knowledge and high-density (HD) constraints coming from general and NS physics:
\begin{eqnarray}
p_{\rm post} (\vec X) = \mathcal{N} \, w_{\rm LD}(\vec X) w_{\rm HD}(\vec X) \, e^{-\chi^2(\vec X)/2} \, p_{\rm prior}(\vec X)  ,
\label{eq:probalikely}
\end{eqnarray}
where $\mathcal{N}$ is the normalisation.
A first filter $w_{\rm LD}$ on the bulk parameters is given by the uncertainty band of the chiral N$^3$LO effective field theory (EFT) calculation for symmetric matter and pure neutron matter by ~\citet{Drischler2016}, which is interpreted as a $90\%$ confidence interval. Moreover, the quality of reproduction  of the experimental nuclear  masses $M_{\rm exp}$ is defined from the error estimator $\chi^2(\vec X,\sigma_0,\sigma_{0,c},b_s,\beta)$:
\begin{equation}
\chi^2= \frac{1}{N_{\rm dof}} \sum_n \frac{\left ( M(A_n,Z_n)-M_{\rm exp}(A_n,Z_n)\right )^2}{\sigma_{n}^2} \ ,
\label{eq:chi2}
\end{equation}
where $N_{\rm dof}$ is the number of degrees of freedom, the sum runs over the AME2016 nuclear mass table from \citet{AME2016}, $M$ is calculated from Eq.~(\ref{eq:mass}) for each model described by a combined set $\{\vec X,\sigma_0,\sigma_{0,c},b_s,\beta\}$, and $\sigma_n$ corresponds to the systematic theoretical error.
As for the $p$ parameter, unless specified otherwise (see Sect.~\ref{sect:surface}), its value was allowed to vary in the range from 2 to 4.

A further constraint is given by the condition that the minimisation of the thermodynamic potential Eq.~(\ref{eq:auxiliary}) leads to physically meaningful results for the crust, namely positive values for the optimal gas and cluster densities.
Finally, the high-density behaviour of the equation of state is also controlled through the $w_{\rm HD}$ filter in Eq.~(\ref{eq:probalikely}) by imposing (i) stability; (ii) causality; (iii) a positive symmetry energy at all densities; and (iv) the resulting equation of state to support $M_{\rm max} > 1.97M_\odot$,  where $M_{\rm max}$ is the maximum NS mass at equilibrium determined from the solution of the Tolmann-Oppenheimer-Volkoff (TOV) equations \citep{hpy2007}, $M_\odot$ being the solar mass.
The resulting posterior equations of state were shown by \citet{Carreau_prc} and \citet{Carreau_these} to be fully compatible with the the recent measurement of the tidal polarisability parameter $\Lambda$ in the gravitational-wave event GW170817 \citep{Abbott2018} (see \citet{Carreau_prc} for details and their Fig.~1).

\begin{figure*}[!htbp]
\begin{center}
\includegraphics[width=0.5\linewidth]{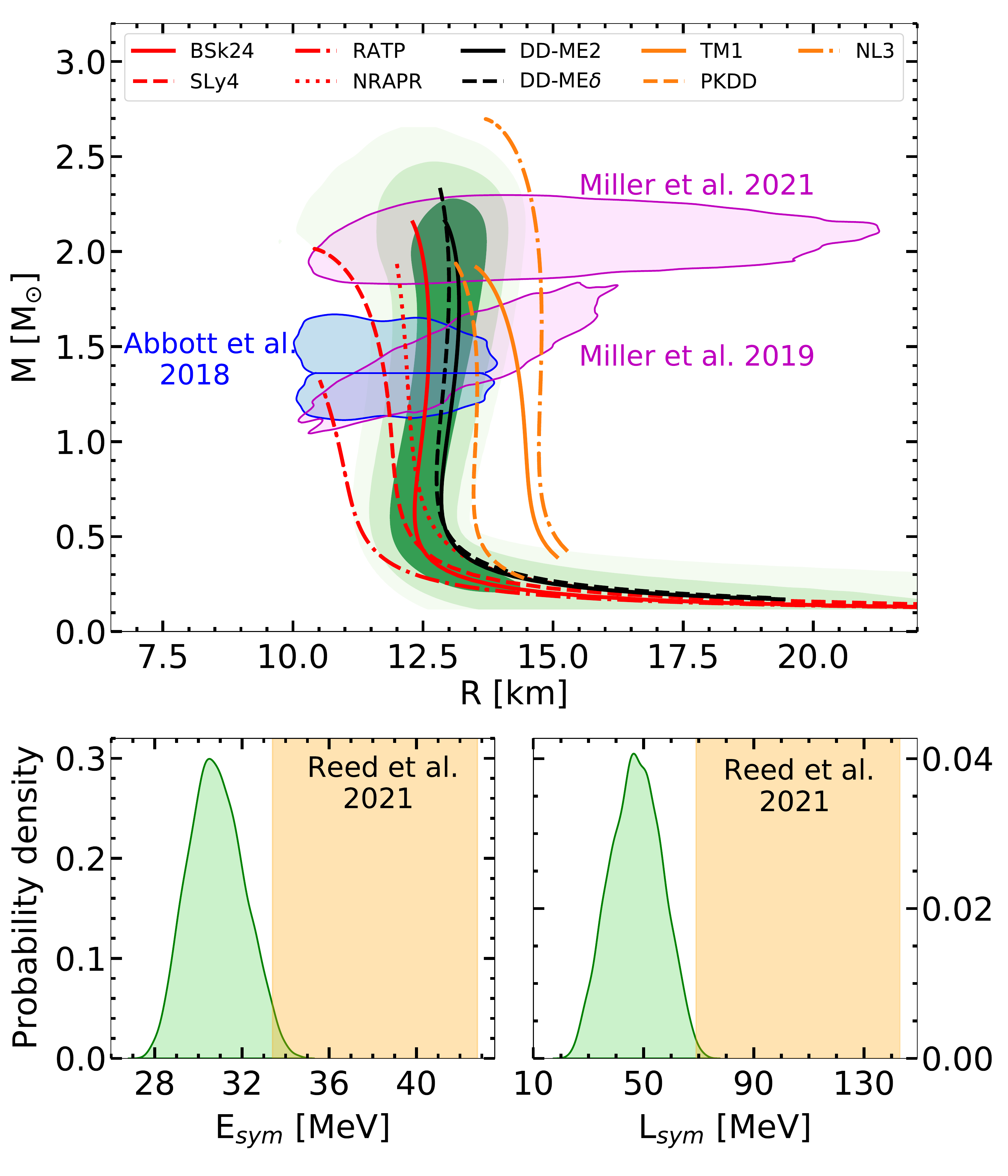}
\end{center}
\caption{Top panel: Marginalised posterior distribution for the neutron-star gravitational mass as a function of radius for the models including both the low- and high-density filters.
The shaded areas correspond to the $1\sigma$, $2\sigma$, and $3\sigma$ (from dark to light green) confidence intervals, while lines show the mass-radius relation for different selected models.
The pink (blue) shaded areas represent the constraint from \citet{Miller2019, Miller2021} (\citet{Abbott2018}) at $2\sigma$.
Bottom panels: Marginalised posterior distributions for the symmetry energy at saturation, $E_{\rm sym}$ (left panel), and its slope, $L_{\rm sym}$ (right panel).
The shaded rectangles show the values of $E_{\rm sym}$ and $L_{\rm sym}$ from \citet{Reed2021}
(see text for details). 
}
\label{fig:MR-sym}
\end{figure*}

Out of the $10^8$ models generated to numerically sample the prior parameter distribution, 7008 models are retained by the applied filters\footnote{In order to have comparable statistics, when the low-density EFT constraint is applied from 0.1~fm$^{-3}$ instead of 0.02~fm$^{-3}$, $2 \times 10^6$ models are generated, of which 7714 are retained. Moreover, in the Bayesian analysis, to speed up the computation, the composition of the different phases are fixed to those found for the spheres. We checked that this assumption does not change the sphere-pasta transition point considerably.} to compute marginalised posteriors for the different observables $Y$ whose average value is given by
\begin{equation}
\langle Y \rangle = \prod_{k=1}^{N} \int_{X_k^{\rm min}}^{X_k^{\rm max}} dX_k Y(\vec X) p_{\rm post}(\vec X),
\label{eq:distri}
\end{equation}
where $p_{\rm post}(\vec X)$ is the posterior distribution, $Y(\vec X)$ is the value of the $Y$ variable as obtained with the 
$\vec X$ parameter set, $X_k^{\rm min(max)}$ is the minimum (maximum) value in the prior distribution of parameter $X_k$ taken from Table~2.2 in \citet{Carreau_these} (see also \citet{Margueron2018a} for details)\footnote{At variance with Table~2.2 in \citet{Carreau_these}, which reports the minimum (maximum) value of the $b$ parameter as 1 (10), here we vary the $b$ parameter in the $[1,6]$ range. We checked that the latter range is sufficient to satisfactorily explore the equation-of-state space.} and of $p$ from 2 to 4, and $N=14$ is the number of parameters defining the nuclear model. 
The marginalised posteriors of given observables can then also be confronted with additional constraints, not originally included in the filters, to check the validity of the assumptions and the flexibility of the prior.
An example of such comparison with both recent astrophysical and nuclear-physics data is given in Fig.~\ref{fig:MR-sym}.
In the top panel, we show the marginalised posterior for the gravitational mass of the NS as a function of the radius.
The shaded green areas correspond to the 68\% ($1\sigma$), 95\% ($2\sigma$), and 99\% ($3\sigma$) (from dark to light green) confidence intervals, while the blue areas represent the constraint inferred from the gravitational-wave event GW170817 by the LIGO/Virgo collaboration \citep{Abbott2018}, and the pink areas show the recent results from NICER \citep{Miller2019, Miller2021}, all at the $2\sigma$ level.
For comparison, in the same figure we also show the mass-radius relation obtained with the nine models analysed in Sect.~\ref{sect:models-cf}.
We can see that our posterior distribution is in agreement with the recent observations coming from NICER \citep{Miller2019}, NICER combined with XMM-Newton \citep{Miller2021}, and those inferred from GW170817 \citep{Abbott2018}.
On the other hand, some of the popular models displayed lie outside the posterior distribution, meaning that they would be filtered out by our low- and/or high-density constraints.
In the bottom panels of Fig.~\ref{fig:MR-sym}, we plot the marginalised posterior distributions for the symmetry energy parameters $E_{\rm sym}$ and $L_{\rm sym}$.
Very recently, \citet{Reed2021} inferred the values of $E_{\rm sym}= (38.1 \pm 4.7)$~MeV and $L_{\rm sym}=(106 \pm 37)$~MeV from measurements of the neutron-skin thickness in $^{208}$Pb performed by the PREX collaboration (shown as shaded rectangles in the figure).
These findings point towards rather high values of these coefficients, overestimating the present limits deduced from both theoretical and experimental measurements (see e.g. \citet{Baldo2016, Burgio2018} for a discussion, and references therein).
Indeed, our posterior distributions suggest $E_{\rm sym} = (30.8 \pm 1.3)$~MeV and $L_{\rm sym} = (47.3 \pm 9.2)$~MeV, meaning that the average values we obtain are considerably lower than those reported in \citet{Reed2021}.
However, the estimation of these parameters made in the latter work is not model independent, and it is based on a correlation extracted by employing a limited number of models that might not span all the possible functional dependences.  
In our study, the low-density constraint imposed by EFT calculations would disfavour such high values of the symmetry energy parameters, and models that predict high $E_{\rm sym}$ and $L_{\rm sym}$ are filtered out in our posterior distribution.
This is in full agreement with \citet{Reed2021}, where the tension between the extracted values of $L_{\rm sym}$ and the ab initio calculations was already highlighted (see Fig.~1 of \citet{Reed2021}). 
We can also observe that the NL3 and TM1 models that have, among the models considered in Sect.~\ref{sect:models-cf}, the closest $E_{\rm sym}$ and $L_{\rm sym}$ values to those reported in \citet{Reed2021}, lie outside our posterior distribution and appear in disagreement with the recent astrophysical constraints from GW170817 (see top panel of Fig.~\ref{fig:MR-sym}).

The theoretical ab initio chiral EFT calculation and the experimental AME2016 mass measurements constitute two powerful and complementary constraints allowing us to predict pasta observables with controlled uncertainties. The theoretical filter only acts on the bulk parameters, and it is particularly effective in constraining the parameters in the symmetry sector $\vec X_{\rm sym}$. Conversely, the mass filter is not very effective in selecting the bulk parameters, because a stronger binding from the bulk term can be compensated by an increased importance of the surface terms via the correlation imposed by Eq.~(\ref{eq:mass}). However, the mass filter is crucial in fixing the surface properties of inhomogeneous matter and, consequently, the crustal properties of NSs. 
These two aspects are analysed separately in the next sub-sections.

%%%%%%%%%
\subsection{The influence of the bulk functional}
\label{sect:LD}
%%%%%%%%

\begin{figure*}[!htbp]
\begin{center}
\includegraphics[width=0.8\linewidth]{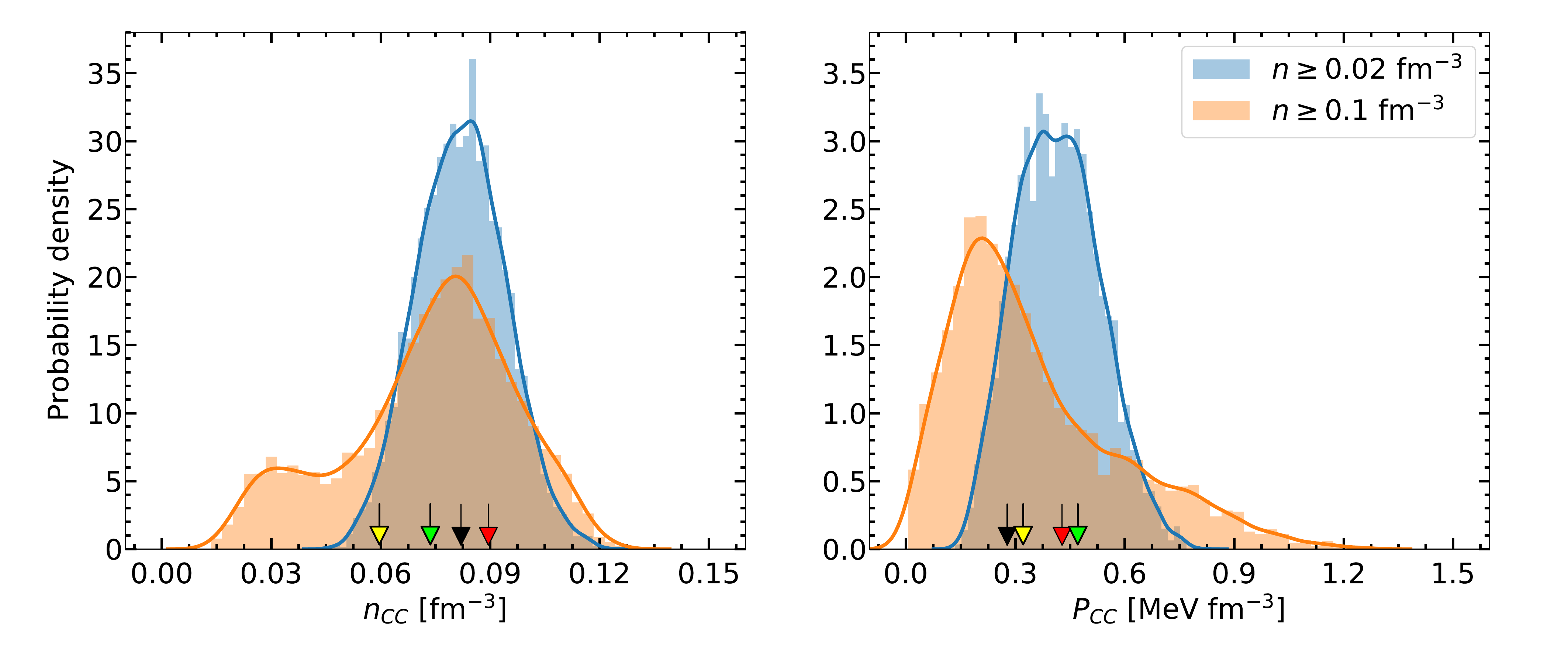}
\end{center}
\vspace{-0.5cm}
\begin{center}
\includegraphics[width=0.8\linewidth]{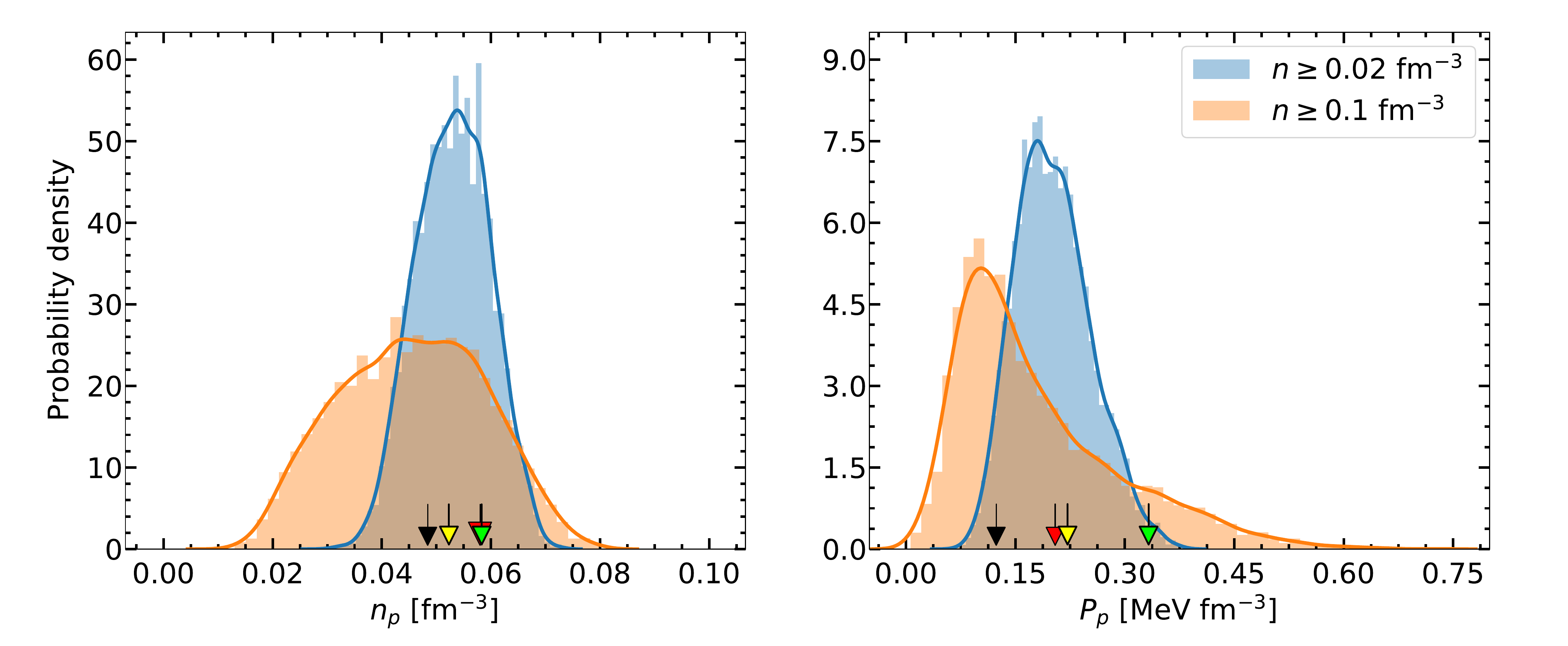}
\end{center}
\vspace{-0.5cm}
\caption{Posterior distributions of crust-core (top panels) and sphere-pasta (lower panels) transition density (left panels) and pressure (right panels), obtained considering different density intervals for the application of the chiral EFT constraint.
Arrows correspond to the predictions of some selected models: BSk24 (black), RATP (red), DD-ME2 (green), and TM1 (yellow)
(see text for details). 
}
\label{fig:cc_pp}
\end{figure*}

\begin{figure}[!htbp]
\begin{center}
\includegraphics[width=0.8\linewidth]{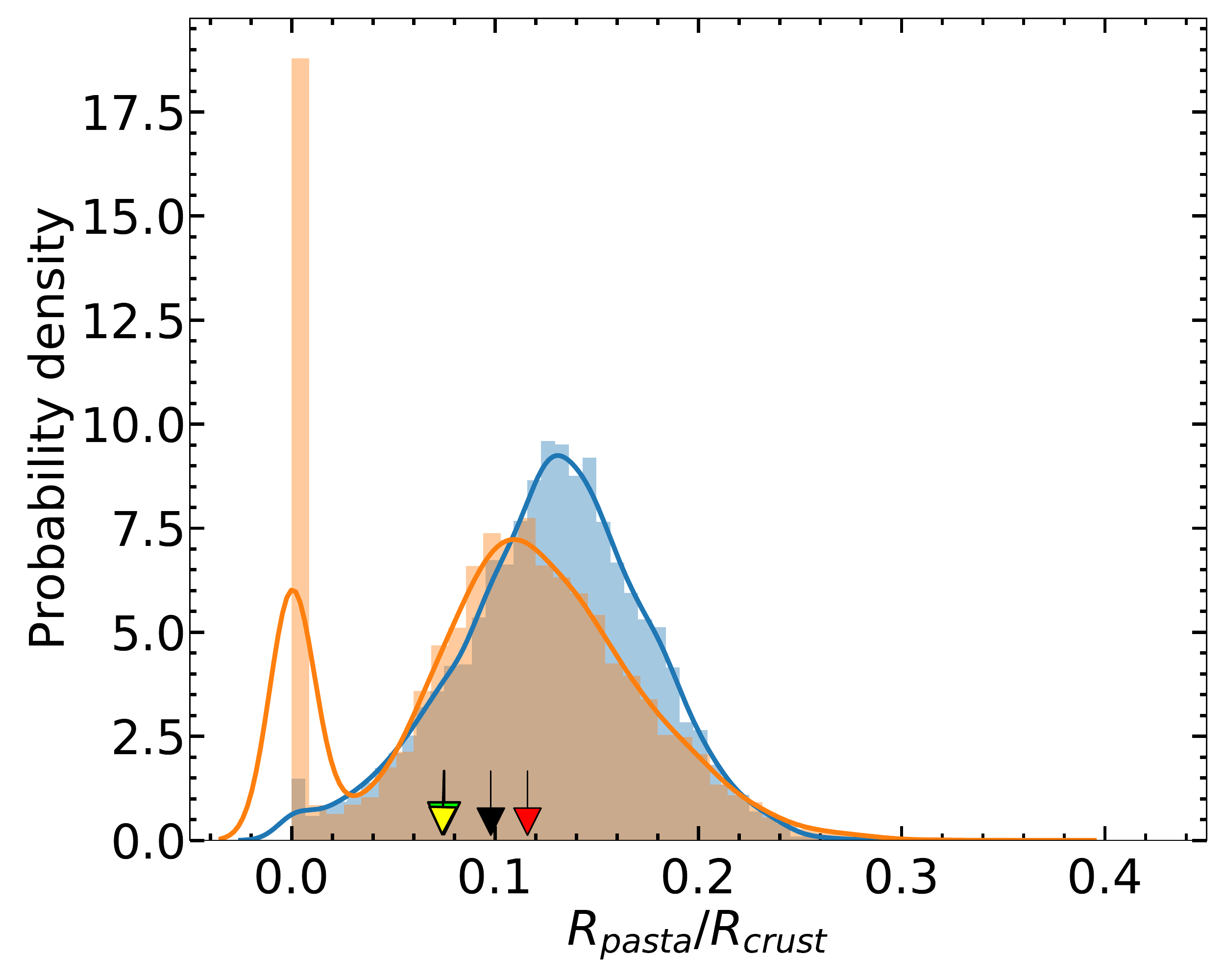}
\includegraphics[width=0.8\linewidth]{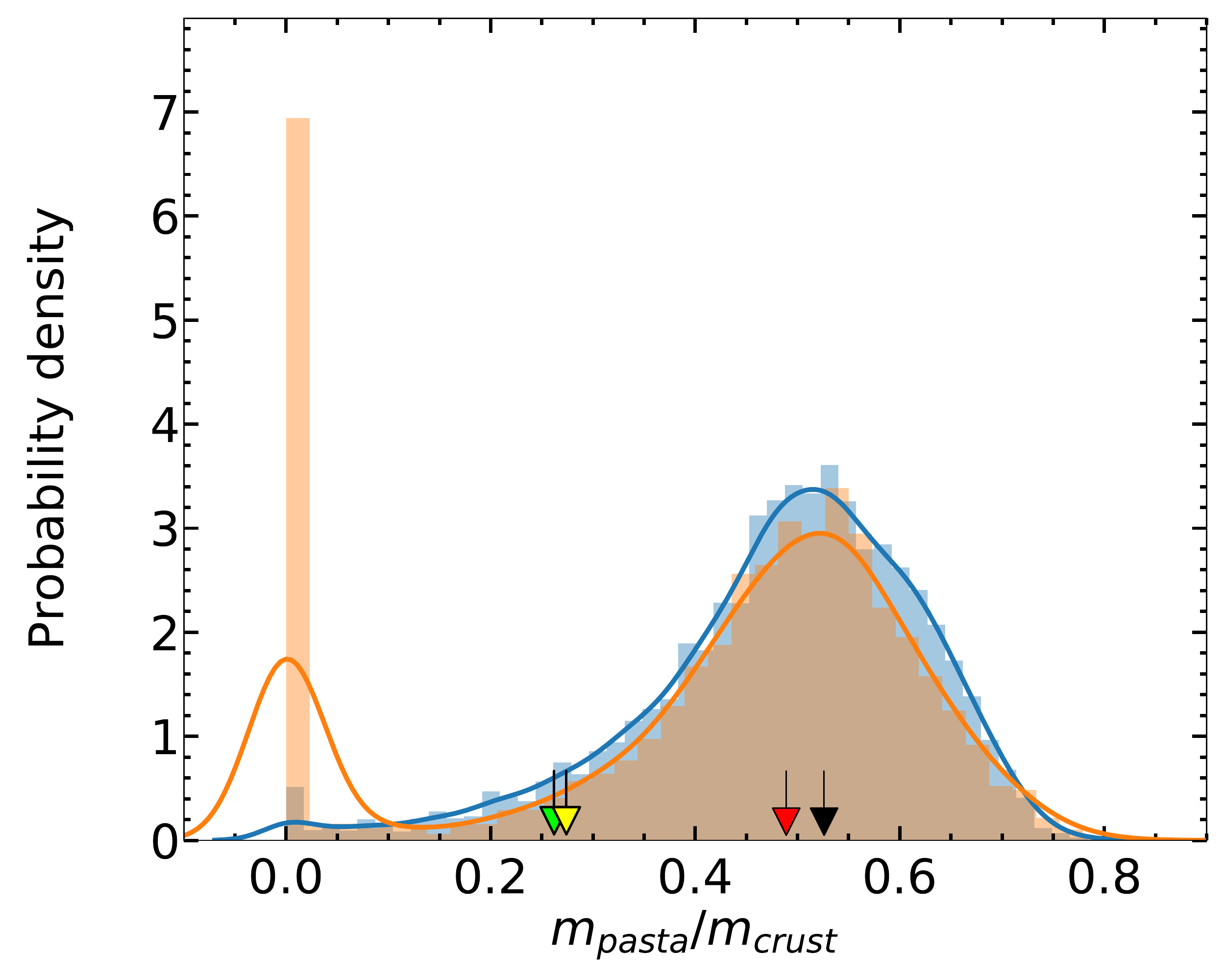}
\includegraphics[width=0.8\linewidth]{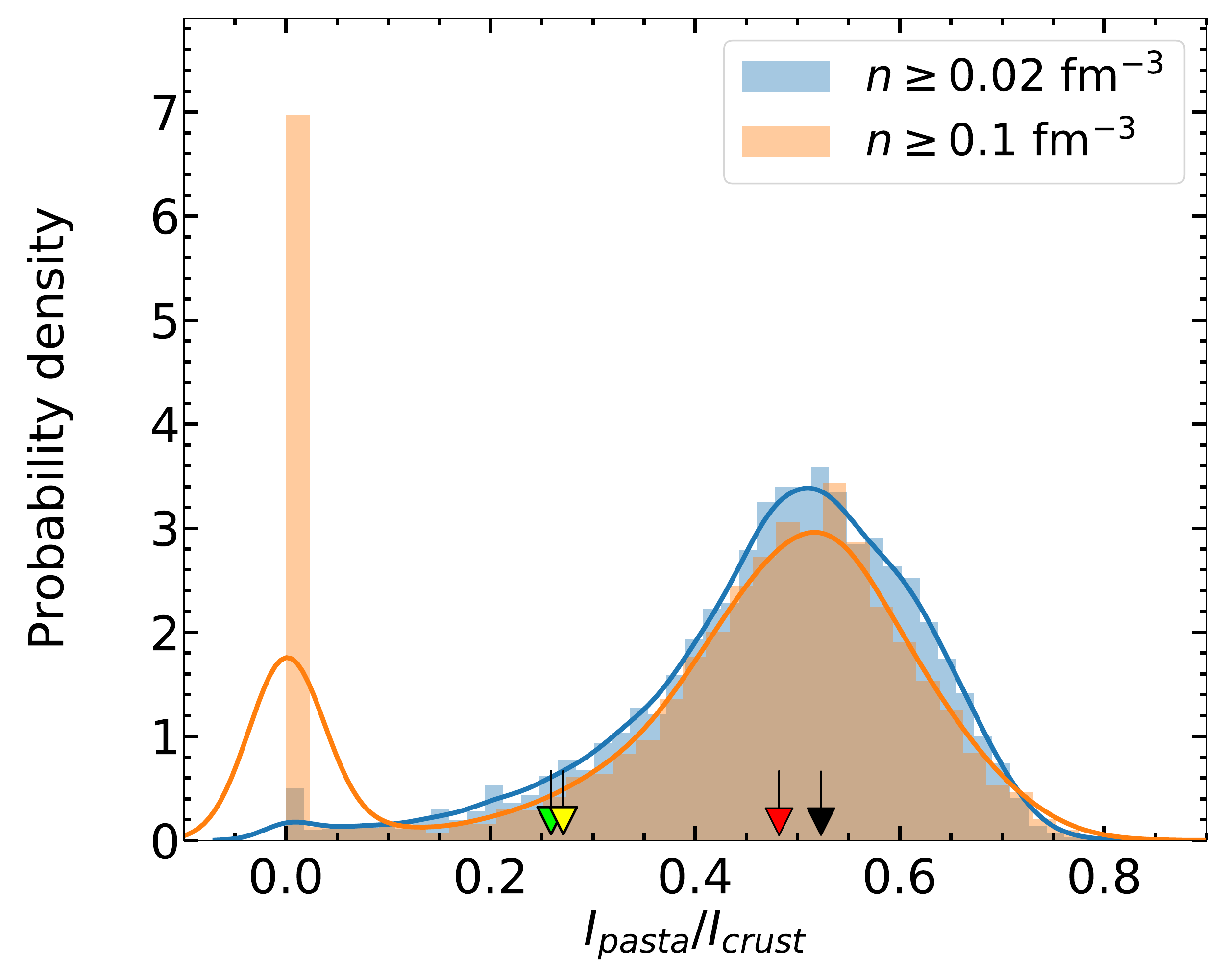}
\end{center}
\caption{Posterior distributions of the thickness (top panel), mass (middle panel), and fractional moment of inertia (lower panel) of the pasta layer with respect to the whole crust for a neutron star with a mass $M = M_{\rm max}$.
The low-density EFT constraints and the arrows corresponding to the predictions of selected models are the same as in Fig.~\ref{fig:cc_pp} (see text for details).
 }
 \label{fig:observables_LD}
\end{figure}

The chiral EFT constraint  was applied in numerous previous studies of static properties of NSs (\citet{Carreau_prc,Lim2019,Tews2019,Essick2020,Brown2020}). 
In these studies, the accent is typically put on the high-density equation of state, which is the dominant ingredient of the different astrophysical observables that are integrated over the whole NS volume, such as the tidal polarisability or the NS mass and radius. 
The compatibility of the functionals with the ab initio predictions in the very low-density region below $n\approx 0.05$ fm$^{-3}$ has a negligible influence for these global NS observables, and as such it was often overlooked.

However, as long as crustal observables are concerned, this very low-density region is crucial. This can be seen in Fig.~\ref{fig:cc_pp}, where we represent the posterior distributions of the density and pressure at the crust-core transition, ($n_{\rm CC},P_{\rm CC}$) (top panels), and the density and pressure at the transition from spherical nuclei to non-spherical pasta structures, ($n_{p},P_{p}$) (lower panels). 
For comparison, we also indicate the predictions corresponding to some selected models with arrows: BSk24 (black-filled arrow), RATP (red-filled arrow), DD-ME2 (green-filled arrow), and TM1 (yellow-filled arrow).
Two different density intervals are considered for the application of the chiral EFT filter, namely $[0.1,0.2]$ fm$^{-3}$ and $[0.02,0.2]$ fm$^{-3}$. 
We can see that the consideration of the very low-density part of the equation of state has a notable effect on the determination of the transition points. 
In particular, a significant number of models that correspond to reasonable properties of nuclear matter close to saturation, and therefore fulfill the chiral EFT condition for $n\ge 0.1$ fm$^{-3}$, are seen to produce very low values for the crust-core and the sphere-pasta transition point. 
This implies a very thin crust for the NS and a small or negligible contribution of the pasta phases, as shown in Fig.~\ref{fig:observables_LD}, which displays the prediction for the relative crustal thickness associated with the non-spherical pasta structures, $R_{\rm pasta}/R_{\rm crust}$ (top panel), the associated mass, $m_{\rm pasta}/m_{\rm crust}$ (middle panel), and the fraction of the moment of inertia, $I_{\rm pasta}/I_{\rm crust}$ (lower panel).
The two choices for the low-density constraints and the arrows indicating the predictions of some selected models (BSk24, RATP, DD-ME2, and TM1) are the same as those adopted in Fig.~\ref{fig:cc_pp}. 
These observables are computed from the numerical solution of the TOV equations, the crust radius (mass) being calculated as the difference between the total NS radius (mass) and that of the core (we considered NSs with a total mass equal to the maximum mass predicted by each computed model). 
As can be seen in Figs.~\ref{fig:cc_pp} and \ref{fig:observables_LD}, when the chiral EFT constraint is applied from $n\ge 0.02$~fm$^{-3}$, almost all models corresponding to a very small (or even null) pasta contribution are filtered out of the posterior distribution. 
Knowing that the precision and reliability of the ab initio calculations improve with decreasing density, we can conclude that our present theoretical understanding of nuclear matter implies that pasta phases should exist in the inner crust of NSs.
 
\begin{figure}[!htbp]
\begin{center}
 \includegraphics[width=1.0\linewidth]{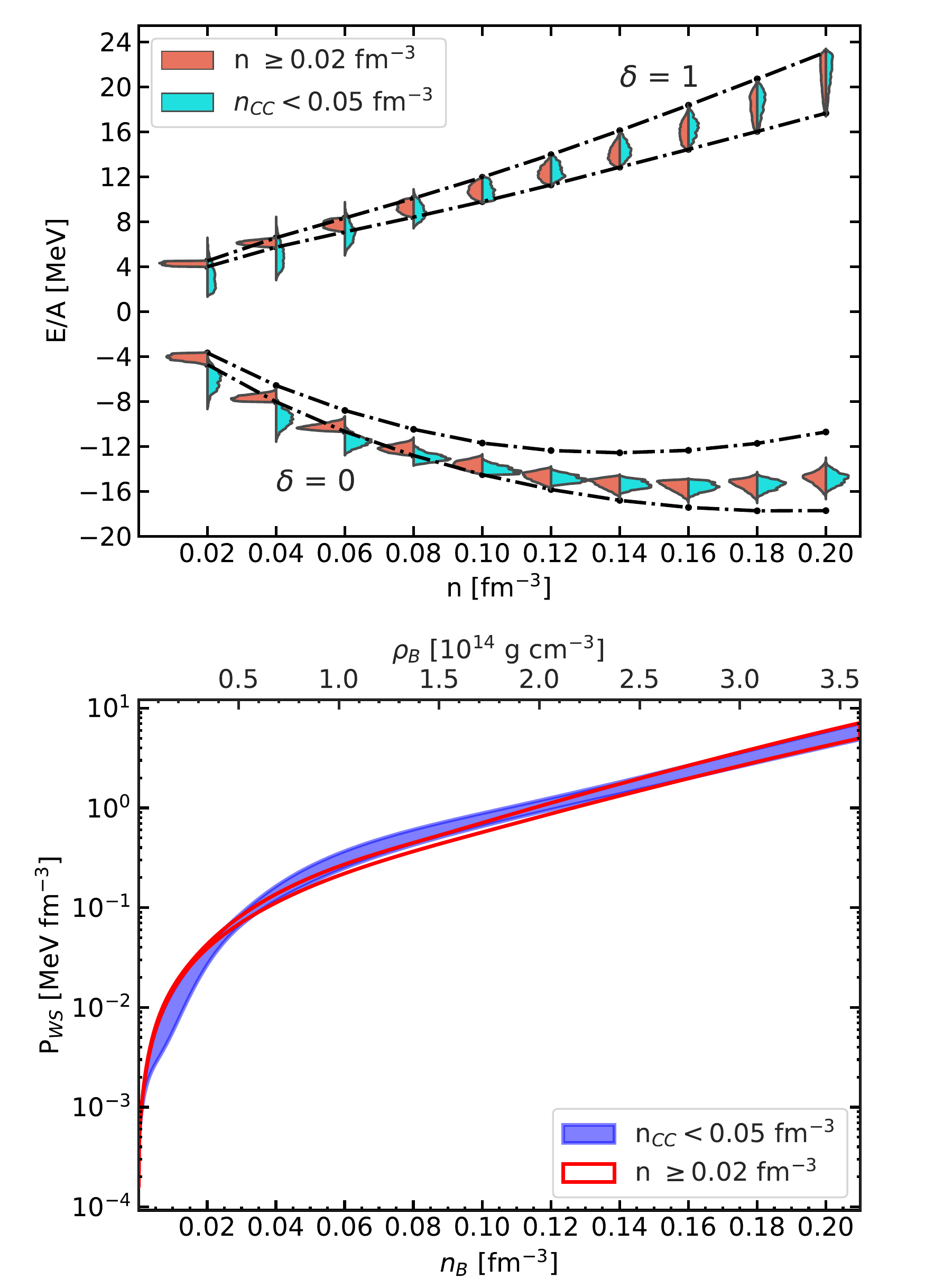}
\end{center} 
%\vspace{-1cm}
\caption{Top panel: Bands of the energy per baryon of symmetric ($\delta=0$) and pure neutron matter ($\delta=1$) as a function of density representing the chiral EFT constraint from \citet{Drischler2016}. The probability distributions of models for which EFT constraints are applied from $n\ge 0.02$~fm$^{-3}$ ($n\ge 0.1$~fm$^{-3}$ and predicting a crust-core transition $n_{\rm CC} < 0.05$~fm$^{-3}$) are represented as a violin-like shape.
Lower panel: $1\sigma$ band of the total pressure versus baryon density $n_B$ and versus mass-energy density $\rho_B$ (in units of $10^{14}$~g~cm$^{-3}$) predicted by these models (see text for details).}
\label{fig:eos}
\end{figure}

To investigate this point further, in the top panel of Fig.~\ref{fig:eos} we plot the energy per baryon distribution of the different models for several baryon densities, for symmetric ($\delta=0$) and pure neutron matter ($\delta=1$). The corresponding uncertainty bands of the chiral EFT constraint from \citet{Drischler2016} are also given (delimited by black dash-dotted lines).
Models for which EFT constraints are applied from $n\ge 0.02$~fm$^{-3}$ ($n\ge 0.1$~fm$^{-3}$ but predicting a crust-core transition lower than $0.05$~fm$^{-3}$, see Fig.~\ref{fig:cc_pp}), are displayed in coral (light blue) on the left (right) part of the density axis.
We can clearly see that models that are not compatible with the EFT constraints in the range $0.02 \le n < 0.1$~fm$^{-3}$ and yielding a crust-core transition below $0.05$~fm$^{-3}$ also predict lower energy per baryon in the very low-density region. 
This results in a stiffer equation of state in the NS crust in the baryon density range $0.03 \lesssim n_B \lesssim 0.1$~fm$^{-3}$ (or, equivalently, in the mass-energy range $0.5 \times 10^{14} \lesssim \rho_B \lesssim 1.8 \times 10^{14}$~g~cm$^{-3}$), as can been seen by comparing, in the lower panel of Fig.~\ref{fig:eos}, the red and blue bands that represent the total pressure in the Wigner-Seitz cell as a function of the baryon (lower axis) and mass-energy (top axis) density in the inner crust.
It is also clear that applying the EFT constraint from $n\ge 0.02$~fm$^{-3}$ induces a stricter filter and thus a narrower band for the resulting equation of state.
Figures~\ref{fig:cc_pp}, \ref{fig:observables_LD}, and \ref{fig:eos}  thus show that it is essential to verify the compatibility with the ab initio calculations down to very low densities to obtain realistic predictions for the properties of the NS crust.
Incidentally, \citet{Shelley2021}, who performed a systematic study of the NS inner-crust composition using several Skyrme functionals, also pointed out the importance of constraining the low-density part of the equation of state.

\begin{figure*}[!htbp]
\begin{center}
\includegraphics[width=\linewidth]{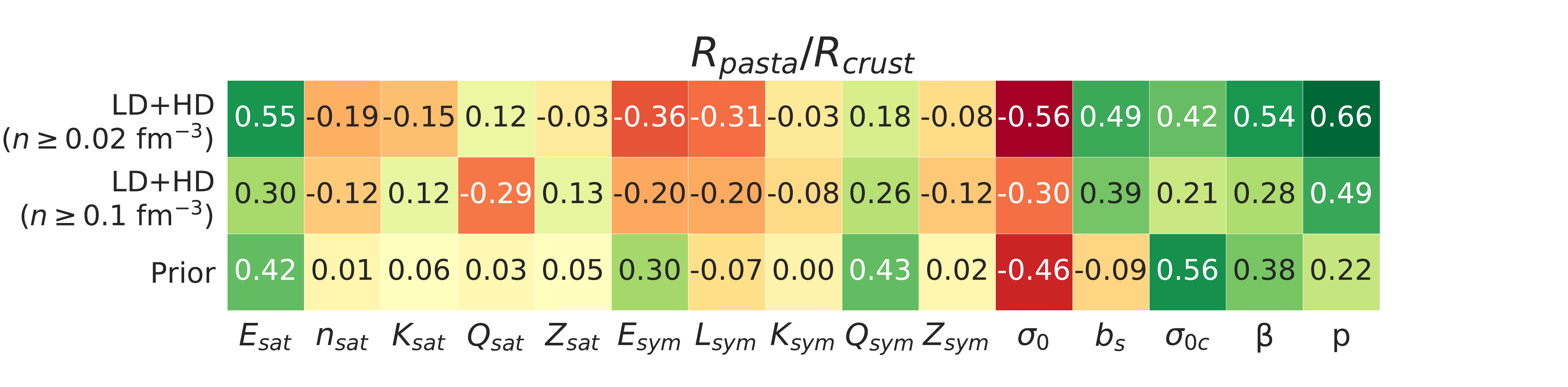}
\includegraphics[width=\linewidth]{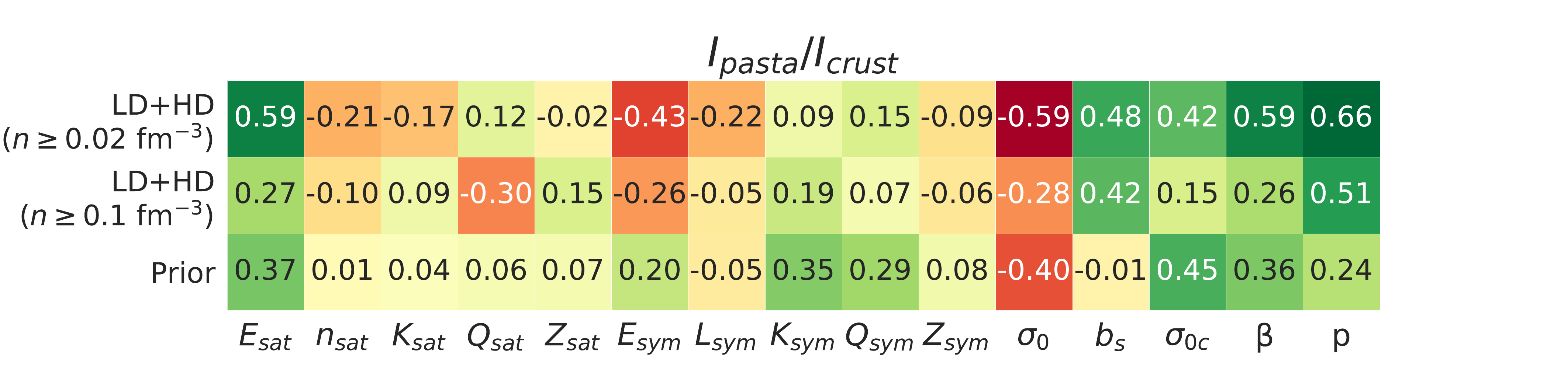}
\end{center} 
%\vspace{-1cm}
\caption{Pearson correlations between pasta properties and the bulk, surface, and curvature parameters.
Two different density intervals for the application of the chiral EFT constraint are considered (see text for details). }
\label{fig:corr}
\end{figure*}

The relative influence on the pasta observables of the different parameters governing the bulk properties of nuclear matter can be quantified from the Pearson linear correlation coefficients displayed in Fig.~\ref{fig:corr} for the pasta thickness with respect to the crustal one (top panel), and the relative fractional moment of inertia (lower panel).
We do not display the Pearson coefficients for the mass of the pasta layers here since they are equivalent to those of the fractional moment of inertia.
In the absence of physical constraints (line labelled `prior' in Fig.~\ref{fig:corr}), no very strong correlation can be observed between the bulk parameters and the pasta observables, as expected from the even exploration of the parameter space of our prior distribution.  
Some loose correlation is observed with $E_{\rm sat}$ and the second ($K_{\rm sym}$) and third ($Q_{\rm sym}$) derivatives of the symmetry energy at saturation $n_{\rm sat}$. 
The prior distributions of the fourth-order parameter $Z_{\rm sym}$ and of the high-order parameters in the symmetric energy sector, $Q_{\rm sat}$ and $Z_{\rm sat}$, are also very large. 
However, they are less influential because the inner crust is characterised by a very low proton fraction (see Fig.~\ref{fig:yp}), and its density is close to $n_{\rm sat}$, which implies a small contribution of the high-order derivatives in the Taylor expansion Eq.~(\ref{eq:etotisiv}). 
When only the models compatible with our present knowledge of nuclear matter are retained  (lines labelled `LD+HD' in Fig.~\ref{fig:corr}), the correlations with the physically significant parameters start to appear. 
It is interesting to observe that the energy functional must be controlled in the whole sub-saturation region (line labelled `$n\ge 0.02$ fm$^{-3}$' in Fig.~\ref{fig:corr}) for the correlations to appear. 
In the absence of this constraint, large compensations are possible among the different terms of the functional, thus washing up the physical correlations.
The most relevant parameters are seen to be the energy of symmetric matter $E_{\rm sat}$ and symmetry energy $E_{\rm sym}$ at saturation, as well as the slope of the symmetry energy $L_{\rm sym}$.
These parameters are already relatively well constrained by nuclear theory and experiments, but we can expect that their uncertainty will be further reduced in upcoming studies, which will lead to an increased precision in the determination of the pasta contribution to the physics of the crust.

%

%%%%%%%%%
\subsection{The influence of the surface tension}
\label{sect:surface}
%%%%%%%%

Figure~\ref{fig:corr} also shows that a better modelling of the pasta properties would be obtained if we could better constrain the energy functional close to saturation. 
However, the most striking information of Fig.~\ref{fig:corr} is the fact that the surface parameters are 
more influential than the bulk properties, as shown by the high correlation coefficients. 
A part of the uncertainty on the surface tension can certainly be ascribed to the very simplified model of nuclear masses that we employ, Eq.~(\ref{eq:mass}), based on a spherical CLD approximation without accounting for deformation, shell, and pairing effects. A more conceptual limitation on the determination of those coefficients is also the fact that the information on the nuclear mass  is not enough to fully pin down the surface properties, due to the partial compensation between surface and bulk. 

We can thus conclude that the strong influence 
of the surface parameters underlines the importance of a unified treatment of the equation of state, which should be employed not only in developing specific equation-of-state models, but also in the statistical studies of NS properties. 

To assess the importance of a self-consistent calculation of surface properties from the assumed bulk energy functional in the modelling of the NS crust, we performed two calculations where the surface parameters in the Bayesian analysis are fixed to the values obtained from two different accurate fits of extended Thomas-Fermi calculations of a large pool of nuclei. 
In the work by \citet{Carreau2020}, the full BSk24 extended Thomas-Fermi mass table was fitted from dripline to dripline, while \citet{Furtado2020} used the SLy4 functional and included in the fit extended Thomas-Fermi calculations beyond drip with global proton fractions as low as $Y_p^{\rm WS}=0.02$.
The resulting surface parameters are reported in Table \ref{tab:surfopt}. 
These fits lead to surface energy functionals slightly different from the ones obtained with the parameters of Table \ref{tab:surfpar} because of the larger pool of nuclei included in the optimisation. 
Moreover, in both cases, the extension of the mass table to extremely neutron-rich nuclei has allowed also an optimal determination of the $p$ parameter (see Eq.~(\ref{eq:surface})), which governs the extreme isospin behaviour of the surface tension and cannot be fixed from the properties of terrestrial nuclei.

\begin{table}[!htbp]
   \caption{Surface and curvature parameters optimised to reproduce the full extended Thomas-Fermi mass table obtained from the BSk24 and SLy4 functionals. }
   \centering
  \begin{tabular}{lccccc}
    \toprule
    &  $\sigma_0$ & $b_s$ & $\sigma_{0,c}$ & $\beta$ & $p$ \\
    & (MeV/fm$^2$) & & (MeV/fm) & & \\
    \midrule
    BSk24 &  0.98636 & 36.227 & 0.09008 & 1.1631 & 3.0 \\
    SLy4 &   0.99654 & 49.82 & 0.061768 & $y_p$+1 & 3.4 \\
    \bottomrule
  \end{tabular}
  \tablefoot{Values of the parameters for BSk24 are taken from \citet{Carreau2020}, and those for SLy4 are taken from \citet{Furtado2020}.}
\label{tab:surfopt}
\end{table}

\begin{figure}[!htbp]
\begin{center}
\includegraphics[width=0.8\linewidth]{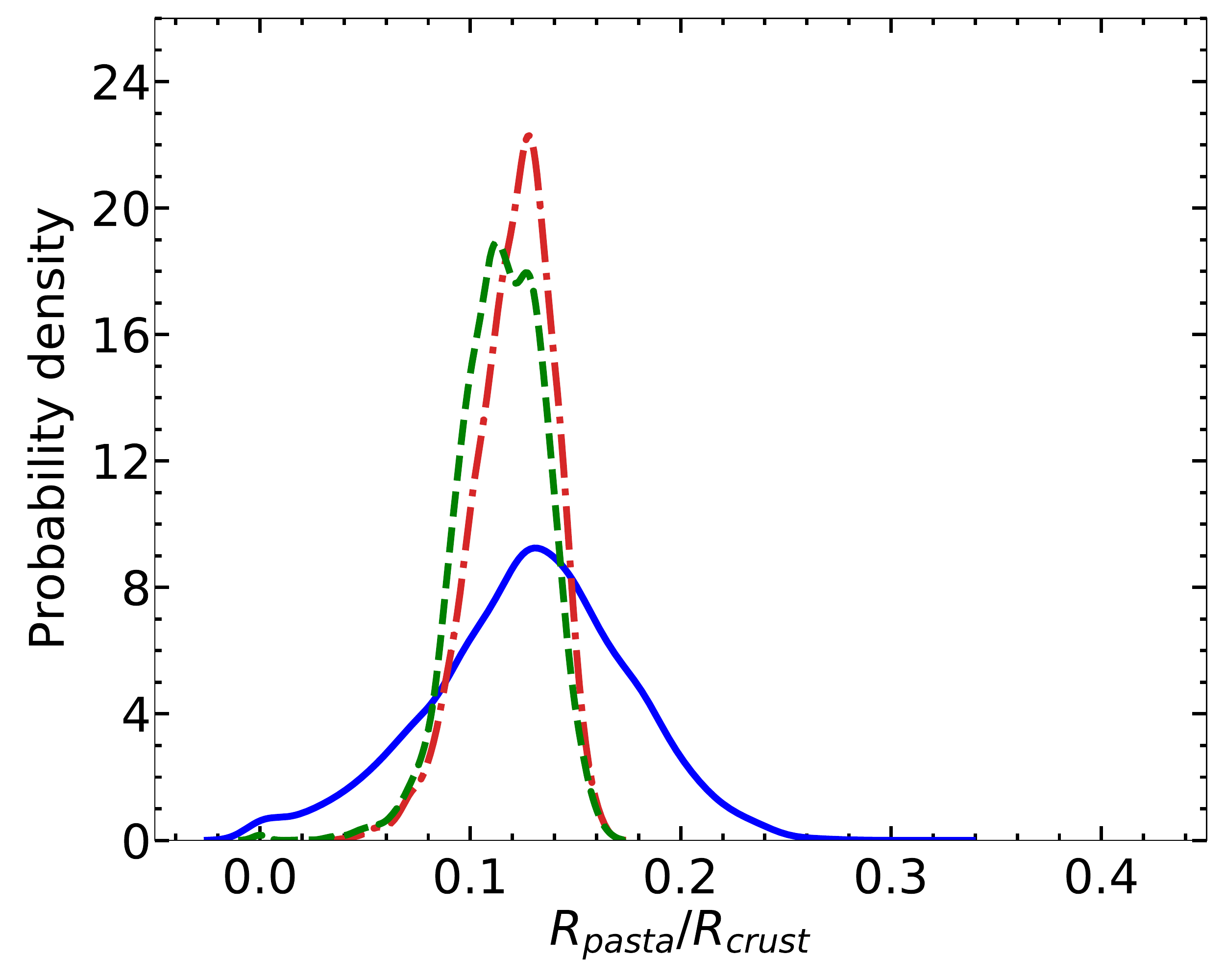}
\includegraphics[width=0.8\linewidth]{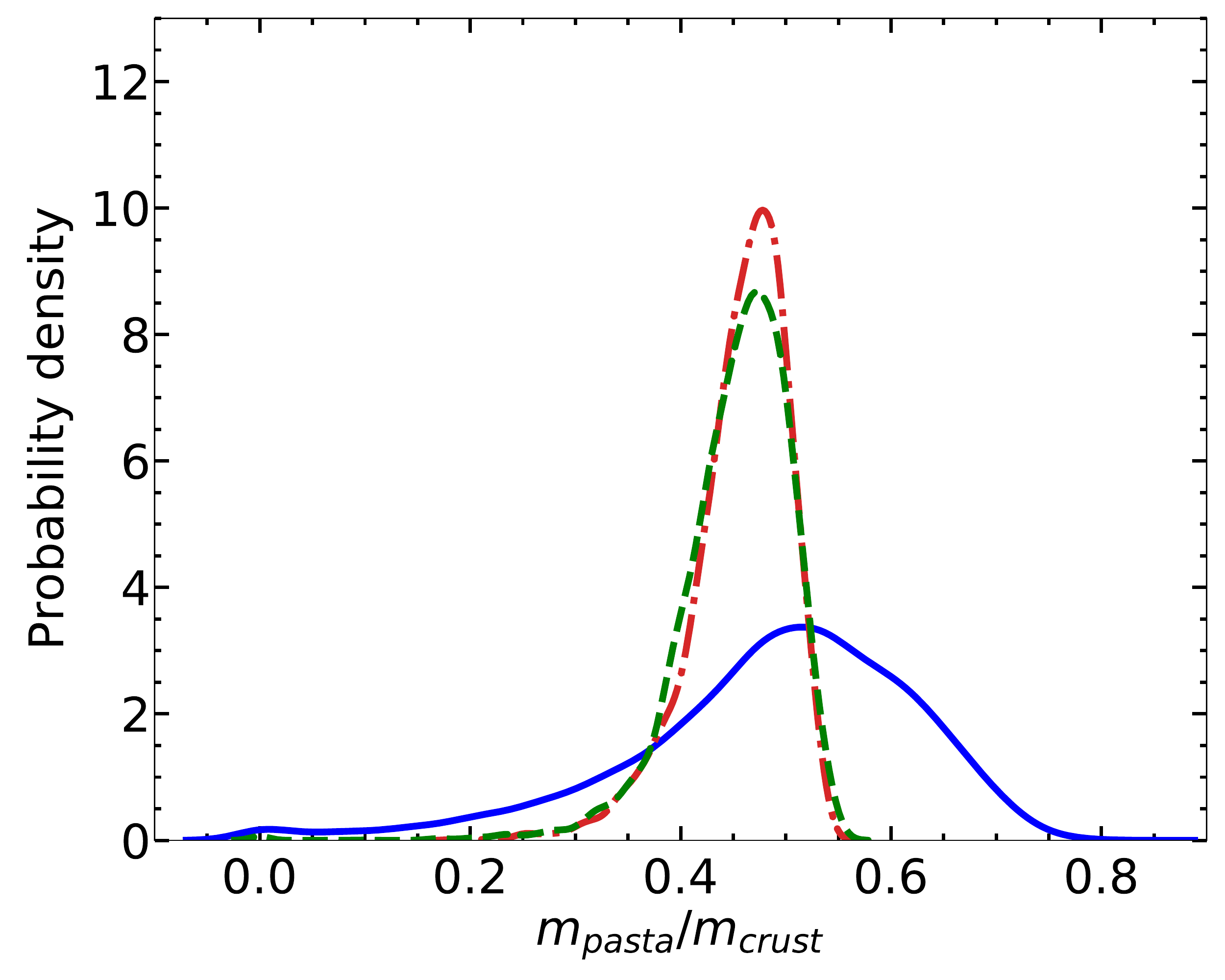}
\includegraphics[width=0.8\linewidth]{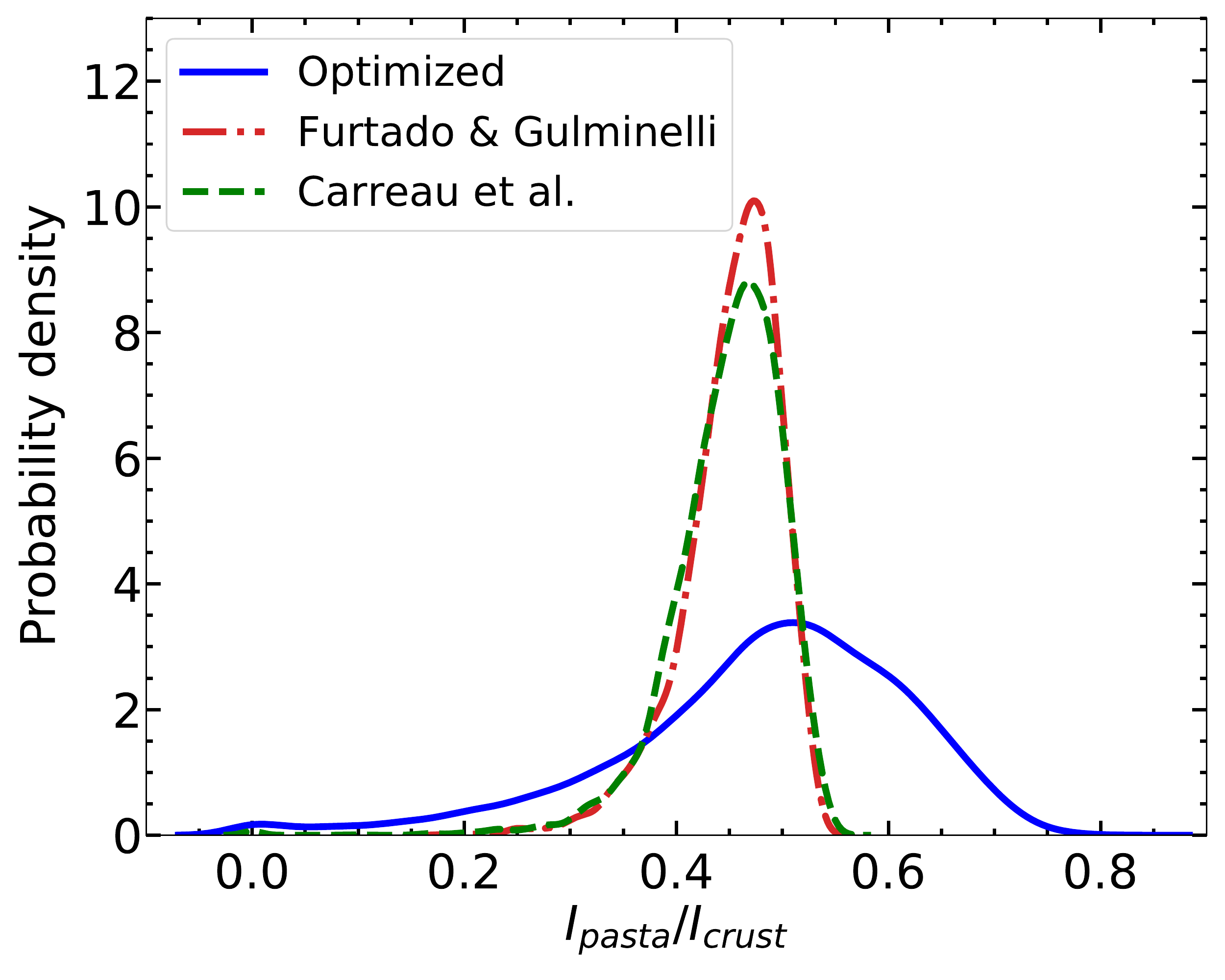}
\end{center}
\caption{Posterior distributions of thickness (top panel), mass (middle panel), and fractional momentum of inertia (bottom panel) of the pasta layer with respect to the whole crust. 
Blue solid curves: surface parameters optimised on the energy functional; red dash-dotted (green dashed) curves: fixed surface parameters from \citet{Furtado2020} (\citet{Carreau2020}). The chiral EFT filter is applied from $n=0.02$ fm$^{-3}$. 
}
\label{fig:observables}
\end{figure}

The surface tensions in Table~\ref{tab:surfopt} are more accurately determined than those deduced from the global reproduction of measured nuclear masses (see Table \ref{tab:surfpar}) and which depend on the empirical $\vec{X}_{\rm bulk}$ parameters.
However, they are not consistent with the bulk properties of functionals different from BSk24 and SLy4, respectively.
The resulting posterior distribution\footnote{In this analysis, the low-density filter is applied from $0.02$~fm$^{-3}$; out of the $10^8$ computed models, 7147 (7024) are retained when the surface parameters are fixed to the values optimised for SLy4 (BSk24).} for the pasta obervables is displayed in Fig.~\ref{fig:observables}. 
We can see that the use of a surface tension that is not consistent with the bulk functional (green dashed lines and red dash-dotted lines) leads to a small shift in the most probable values of the represented pasta-layer properties, and an important deformation of the distribution, with a clear underestimation of the uncertainties in the pasta observables. 

%%%%%%%%%
\section{Conclusions}
\label{sect:conclus}
%%%%%%%%

We presented a study of the properties of pasta phases in cold catalysed NSs, extending the work of \citet{Carreau2019, Carreau2020}.
In particular, we applied a CLD model with surface parameters adjusted either on experimental nuclear masses or theoretical calculations.
We addressed the model dependence of the results by employing different nuclear models, which are reproduced by fixing the bulk parameters to reproduce the low-density behaviour of symmetric matter and pure neutron matter of the corresponding functional.
We find that the value of the transition densities between different geometries is model dependent.
The sequence of the predicted geometries, namely spheres, rods, slab, and tubes, appears to be the same for all considered models, while the presence of bubbles before matter becomes homogeneous in the core is only predicted by some of the considered models.
Moreover, our findings show that the uncertainty in the predictions due to the choice of the nuclear functional is more significant than that due to the many-body method adopted to describe the inhomogeneities.

To assess the model uncertainties in the pasta observables, we performed a Bayesian analysis by largely varying the model parameters using uniform priors and generating posterior distributions with filters accounting for both our present low-density nuclear physics knowledge and high-density general and NS physics constraints.
Our results show that the low-density nuclear physics constraints are crucial in determining the crustal and pasta observables.
Here, the low-density filters, given by the uncertainty band of the chiral N$^3$LO EFT calculation for symmetric matter and pure neutron matter by \citet{Drischler2016} are applied from $n \ge 0.02$~fm$^{-3}$.
Recent works have been devoted to deriving EFT-inspired energy-density functionals benchmarked by ab initio calculations at even lower density ($0 \lesssim n \lesssim 0.1$~fm$^{-3}$), and these could constitute additional constraints in the future \citep{Yang2016}. 
Our final predictions for the different crustal properties, namely the fractional thickness $R_{\rm pasta}/R_{\rm crust}$, moment of inertia $I_{\rm pasta}/I_{\rm crust}$, and mass $m_{\rm pasta}/m_{\rm crust}$ of the pasta layer for a NS with a mass equal to its maximum mass, are summarised in Table~\ref{tab:results}.
The left (right) column lists the results obtained when the low-density EFT filters on the energy of homogeneous matter are applied from $n \ge 0.02$~fm$^{-3}$ ($n \ge 0.1$~fm$^{-3}$).

\begin{table}[htbp]
   \caption{Posterior estimations of the fractional thickness $R_{\rm pasta}/R_{\rm crust}$, moment of inertia $I_{\rm pasta}/I_{\rm crust}$, and mass $m_{\rm pasta}/m_{\rm crust}$ of the pasta layer, normalised to the corresponding crustal quantity.}
  \centering
  \begin{tabular}{lcc}
    \toprule
%    & \;\;\;\;\;\;\;\;\;\;\;\;\;\;\;\;  LD+HD & \\
    & $n\ge 0.02 $ fm$^{-3}$ &  $n\ge 0.1 $ fm$^{-3}$\\
    \midrule
    $R_{\rm pasta}/R_{\rm crust}$               & $0.128\pm 0.047$  & $0.104\pm 0.063$\\ 
    $I_{\rm pasta}/I_{\rm crust}$               & $0.480\pm 0.137$  & $0.411\pm 0.212$\\
    $m_{\rm pasta}/m_{\rm crust}$       & $0.485\pm 0.138$  & $ 0.415\pm 0.214$\\ 
    \bottomrule
  \end{tabular}
  \tablefoot{Values in the table are given for a neutron star with a mass $M = M_{\rm max}$. Both low-density EFT filters and high-density filters are applied. The chiral EFT filter is applied from either $n=0.02$~fm$^{-3}$ (left column) or $n=0.1$~fm$^{-3}$ (right column). The uncertainties indicate $1\sigma$ deviations.}
\label{tab:results}
\end{table}

The most relevant nuclear parameters for predicting the pasta properties are the lower-order empirical parameters, namely the energy of symmetric matter, the symmetry energy, and the slope of the symmetry energy at saturation, as well as the surface and curvature parameters.
In particular, we find that the latter parameters are more influential than the bulk parameters for the description of the pasta observables.
In addition, we show that the use of a surface tension that is inconsistent with the bulk functional leads to an underestimation of the uncertainties in the pasta properties, thus highlighting the importance of a consistent calculation of the nuclear functional.

This formalism can be extended at finite temperature, similarly to the approach adopted in \citet{fantina2020} and \citet{Carreau2020b} for the outer and inner crusts, respectively, to account for the so-called impurities.
This question is of particular interest since the presence of a pasta layer at the bottom of the NS crust may have sizeable impact on different NS phenomena including NS cooling, and this will be addressed in a future work.
Moreover, very recently, upper limits on the NS equatorial ellipticity were provided through gravitational-wave data by the LIGO/Virgo collaboration \citep{Abbott2020}.
The possible impact of pasta phases on this observable, which was already pointed out by \citet{gearheart2011}, thus deserves further investigation.

%%% ACKOWLEDGEMENTS
\begin{acknowledgements}
This work has been partially supported by the IN2P3 Master Project NewMAC, the CNRS PICS07889, and the CNRS International Research Project (IRP) ``Origine des \'el\'ements lourds dans l'univers: Astres Compacts et Nucl\'eosynth\`ese (ACNu)''.
We thank Brynmor Haskell for interesting suggestions.
\end{acknowledgements}

%%%%%%%%%%%%%%%%%%%%%%%
%%% BIBLIOGRAPHY
%\newpage
%\vfill

%%%%%%%%%%%%%%%%%%%%%%%%%%%%%%%%%%%%%%%%%%%%%%%%%%%%%%%%%%%%%%

\end{document}